\documentclass[conference]{IEEEtran}
\IEEEoverridecommandlockouts
\usepackage{cite}
\usepackage{amsmath,amssymb,amsfonts}
\usepackage{algorithmic}
\usepackage{graphicx}
\usepackage{textcomp}
\usepackage{xcolor}

\usepackage[ruled,vlined]{algorithm2e}
\usepackage{makecell}
\usepackage{multirow}
\usepackage{stackengine}
\usepackage{bm}

\usepackage{enumitem}
\usepackage{subfigure}
\usepackage{caption}
\usepackage{url}

\newtheorem{myExample}{Example}
\newtheorem{myDef}{Definition}

\newtheorem{myLemma}{Lemma}

\SetKw{Continue}{continue}
\SetKw{Break}{break}
\SetKwRepeat{Do}{do}{while}

\usepackage{setspace}
\makeatletter
\newcommand\figcaption{\def\@captype{figure}\caption}
\newcommand\tabcaption{\def\@captype{table}\caption}
\makeatother

\def\BibTeX{{\rm B\kern-.05em{\sc i\kern-.025em b}\kern-.08em
		T\kern-.1667em\lower.7ex\hbox{E}\kern-.125emX}}
\begin{document}
	
	\title{Random Walk-based Community Key-members Search over Large Graphs}
	
	\author{
		\IEEEauthorblockN{Yuxiang Wang\textsuperscript{1}, Yuyang Zhao\textsuperscript{1}, Xiaoliang Xu\textsuperscript{1}, Yue Wu\textsuperscript{1}, Tianxing Wu\textsuperscript{2}, Xiangyu Ke\textsuperscript{3}}
		\IEEEauthorblockA{\textsuperscript{1} \textit{Hangzhou Dianzi University, China}, 
			\textsuperscript{2} \textit{Southeast University, Nanjing, China},\\
			\textsuperscript{3} \textit{Zhejiang University, Hangzhou, China}}
		\{lsswyx,yyzhao,xxl,wuyue\}@hdu.edu.cn, tianxingwu@seu.edu.cn, xiangyu.ke@zju.edu.cn
	}
	
	\maketitle
	
\renewcommand{\IEEEQED}{\IEEEQEDopen}
\def\IEEEproofindentspace{2\parindent}
\renewcommand{\IEEEproofindentspace}{10pt}
	
	\begin{abstract}
		Given a graph $G$, a query node $q$, and an integer $k$, community search (CS) seeks a cohesive subgraph (measured by community models such as $k$-core or $k$-truss) from $G$ that contains $q$. It is difficult for ordinary users with less knowledge of graphs' complexity to set an appropriate $k$. Even if we define quite a large $k$, the community size returned by CS is often too large for users to gain much insight about it. Compared against the entire community, key-members in the community appear more valuable than others. To contend with this, we focus on Community Key-members Search problem (CKS). We turn our perspective to the key-members in the community containing $q$ instead of the entire community. To solve CKS problem, we first propose an exact algorithm based on truss decomposition as a baseline. Then, we present four random walk-based optimized algorithms to achieve a trade-off between effectiveness and efficiency, by carefully considering three important cohesiveness features in the design of transition matrix. As a result, we return key-members according to the stationary distribution when random walk converges. We theoretically analyze the rationality of designing the cohesiveness-aware transition matrix for random walk, through Bayesian theory based on Gaussian Mixture Model with Box-Cox Transformation and Copula Function Fitting. Moreover, we propose a lightweight refinement method following an ``expand-replace" manner to further optimize the result with little overhead, and we extend our method for CKS with multiple query nodes. Comprehensive experimental studies on various real-world datasets demonstrate our method's superiority.
	\end{abstract}

	\section{Introduction}
	\label{sec:intro}
	Graphs are the prevalent underlying storage model for many of today's large-scale and real-world information networks \cite{Zhang2019,Fang2017,Huang2017,Sun2020,Liu2020}, e.g., social networks, collaboration networks, and criminal networks. In these graphs, nodes represent entities (e.g., authors in collaboration networks) and edges represent the relationship between two entities (e.g., co-authorship). Community search (CS) on graphs has been studied widely, which is important for valuable communities' exploration and is applied in personalized community analysis \cite{Liu2020,Sozio2010,Xu2022,Wang2023}. Given a graph $G$ and a query node $q$, CS finds a cohesive community from $G$ that contains $q$. In the literature of CS, $k$-core \cite{Cui2014}, $k$-truss \cite{Huang2015}, $k$-ECC \cite{Hu2016}, and $k$-clique \cite{Cui2013} are usually used to measure the community's structure cohesiveness. 
	
	\begin{figure}
		\setlength{\abovecaptionskip}{0.1cm}
		\centering
		\includegraphics[scale=0.3]{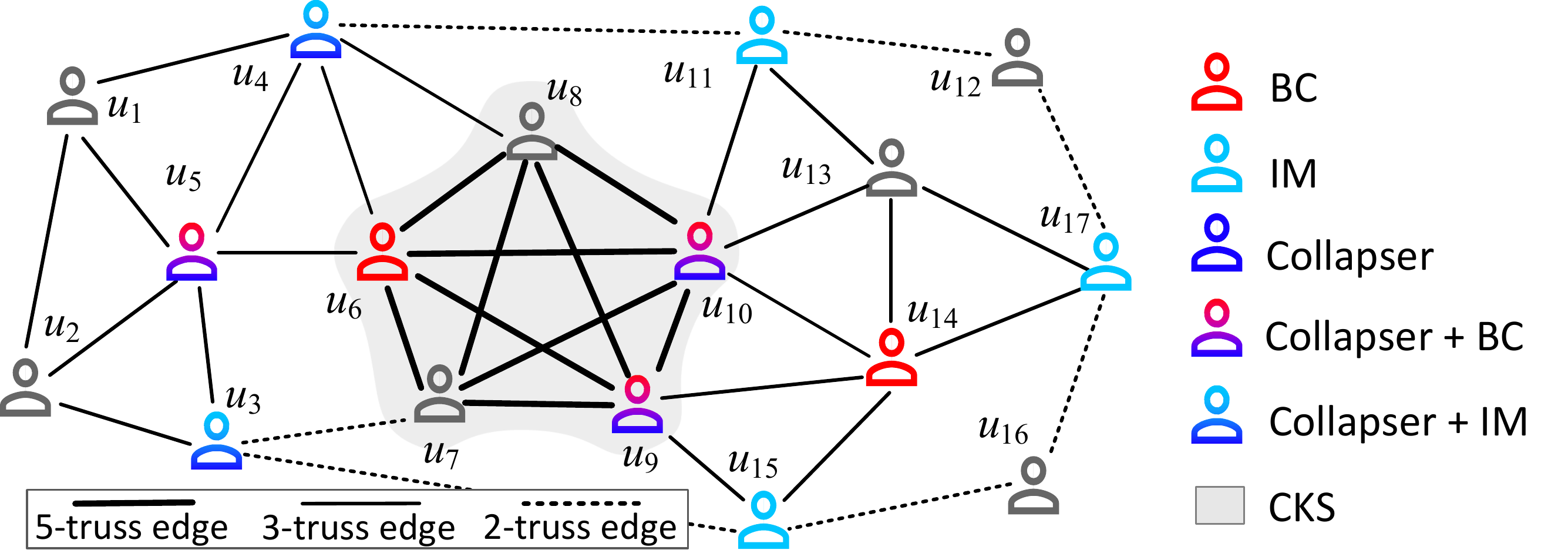}
		\caption{An example of key-members search by using different methods (given the query node as $u_{14}$).}
		\label{fig:example}
		\vspace{-0.5cm}
	\end{figure}
	
	\vspace{0.1cm}
	\noindent\textbf{CKS problem.} In contrast to the classic CS, here we focus on a new problem: the \textbf{C}ommunity \textbf{K}ey-members \textbf{S}earch problem (CKS). This problem stems from the fact that the size of the community returned by CS is often too large for ordinary users to gain much insight about it \cite{Yao2021,Liu2021}. It's also difficult to set an appropriate $k$ for ordinary users with less knowledge of the graphs' complexity (e.g., density, diameter) \cite{Kim2022}. Even if we set a large $k=15$, a community (e.g., measured by $k$-truss) would involve more than 1000 nodes on average for \texttt{Artist} dataset \cite{Rozemberczki2019}, which is a big number for users to analyze. Compared to the entire community, key-members in a community are more valuable \cite{Liu2022,Munikoti2022,Sun2022,Xu2019}. For example, in criminal investigation \cite{Hepenstal2021}, few key-members of a criminal gang are often more valuable than the large number of ordinary gang members. Figure \ref{fig:example} illustrates an example of CKS on a real criminal network consisting of 17 suspects and is gradually sparse from the inside out. Two suspects have an edge if they have a close relationship, e.g., frequent mutual contact, co-occurrences at some places, or close economic dealings. According to \cite{Calderoni2017,Liu2012}, key-members tend to exhibit the greatest structural cohesiveness, such as the middle dense subgraph fromed by $\{u_6,u_7,u_8,u_9,u_{10}\}$ in Figure \ref{fig:example}. Suppose the police only know a marginalized suspect $u_{14}$. If they can leverage $u_{14}$ to find the key-members of the criminal gang to which $u_{14}$ belongs, then an effective strike to this criminal gang would be performed precisely. 
	
	Essentially, key-members in a network usually form a more cohesive subgraph \cite{Calderoni2017,Liu2012}. However, many existing network metrics used for critical nodes identification do not consider the cohesive requirement, yielding different results. Influence is a common metric to measure nodes’ importance \cite{Zhu2020,Zhao2019}, and the influence maximization (IM) aims to find those nodes having the largest influence spread \cite{Wang2017}. In Figure \ref{fig:example}, nodes returned by IM algorithms \cite{Wang2017,Zhu2020} are indicated by cyan-blue ($\{u_3,u_4,u_{11},u_{15},u_{17}\}$). They are spread across the entire network and have weak structure cohesiveness to each other, none of them are located in the middle dense subgraph. Betweenness centrality (BC) is another metric that computes a node's importance in terms of the total number of shortest paths that pass through it \cite{Mumtaz2017}. By applying BC, we obtain nodes indicated by red, still including some less-optimal nodes such as $u_{5},u_{14}$. They are more like middle-level roles of an organization that connect some important nodes (e.g., $u_6$, $u_9$, $u_{10}$) and some ordinary nodes (e.g., $u_1$, $u_{17}$). Moreover, \cite{Zhang2017a} defines a network's important nodes as that will cause a community's collapse from a large $k$-core to a small one if they were removed, called collapsers. Given an integer $k=3$, the collapsers are indicated by blue ($\{u_3$,$u_4$,$u_5$,$u_9$,$u_{10}\}$), still showing weak structure cohesiveness. These definitions are not suit for CKS, because they lack consideration of the close relationship that naturally exists among key-members. Differ from above, in this paper, we apply $k$-truss model to measure a community's cohesiveness and define the key-members of a community as nodes with the maximum trussness, since it is well recognized that $k$-truss has strong structural cohesiveness and high computational efficiency \cite{Liu2020}. Besides $k$-truss, there are still other more cohesive graph models, such as $k$-clique \cite{Cui2013} or ego networks with pre-defined strict density constraints \cite{McAuley2014}. We will extend our solution to them in future. Thus, CKS clearly can be stated as: \textit{given a graph $G$, a query node $q$, we aim to find the key-members with the maximum trussness from the most cohesive community containing $q$} (defined in \S \ref{sec:pre_pro}).
	
	\vspace{0.1cm}
	\noindent\textbf{Applications.} CKS can be applied in many real-world applications. (1) In the field of public security, the police can precisely narrow the scope of investigation and target those key criminals through a small number of known suspects. (2) In the academic area, once we read an inspired paper of a certain researcher, CKS can recommend more high-quality papers of top scholars from the research community containing this certain researcher. (3) In a social network such as Tiktok, users can leverage one of their followings to explore more top-tier vloggers in the specific field they are interested in.
	
	\vspace{0.1cm}
	\noindent\textbf{Our solution.} We first present an exact framework in \textbf{\S \ref{sec:exact_framework}} based on truss decomposition (TD): (1) we find the most cohesive community containing the query node $q$ automatically without a pre-defined $k$ and (2) we identify the key-members with the maximum trussness from the above community. This framework easily adapts to existing representative TD methods, such as TD-bottomup \cite{Wang2012}, TD-topdown \cite{Wang2012}, AccTD \cite{Che2020}, and TCP-Index \cite{Jiang2021,Huang2014}. For simplicity, we briefly introduce exact algorithms based on TD-bottomup and TCP-Index in \textbf{\S \ref{sec:exact_algorithms}} and \textbf{\ref{sec:exact_index}}, respectively. They server as the exact baselines to generate ground truths in our experimental study (\S \ref{sec:exp}). Although the exact algorithms are easy to implement, they are problematic for the following reasons: First, since we do not know the value of $k$, in the worst case, it requires enumerating every possible $k$ to find the maximum $k$-truss $T_k$ in the first step, e.g., enumerating from $k=3$ until no more cohesive $T_k$ with a larger $k$ can be found, which is quite time-consuming. For example, given a dataset with 0.8M edges (M = $10^6$), it requires 17 secs on average. Second, $T_k$'s size in step (1) would significantly affect the efficiency of step (2), e.g., a few of seconds to a dozen seconds depends on $T_k$'s size, and the larger the $T_k$, the slower the step (2). 
	
	In practice, finding key-members may not need a tardy exact result. It is more desirable if a search engine first quickly returns a good enough result, while refining it with an additional lightweight expansion if necessary. This motivates us to present four random walk-based optimized algorithms to achieve a trade-off between effectiveness and efficiency. We aim to find a set of nodes that are most likely to be key-members through random walk on a graph. Intuitively, if we can control a walker towards a node $u$ along a path following the guidance of the community's cohesiveness features, then $u$ is more likely to belong to a $k$-truss with a large $k$, indicating that $u$ would be a key-member with high probability.
	
	\vspace{0.1cm}
	\noindent\underline{(1) Basic random walk-based algorithm (\textbf{\S \ref{sec:rw_basic}}).} Given a graph $G$ and a query node $q$, we design the transition matrix $\bm{P}$ over $G$ based on the support of edge (i.e., the number of triangles that contains an edge). Here, we use an edge's support as the fundamental cohesiveness feature. Then, we conduct random walk based on $\bm{P}$ until it converges. Finally, we return top-$n$ key-members according to the stationary distribution of all nodes in $G$, where each node has a stationary visiting probability showing it's likelihood of belonging to a large $k$-truss. We also present a relaxed version of this algorithm to improve the random walk's efficiency.
	
	
	\vspace{0.1cm}
	\noindent\underline{(2) Optimization with average support (\textbf{\S \ref{sec:rw_op1}}).} We optimize the basic algorithm's effectiveness through a refined transition matrix $\bm{P}$. Given two nodes $u,v$ linked by an edge $e_{uv}$, we expect to move from $u$ to $v$ with a large transition probability $p_{uv}$ when $v$ is likely to belong to a more cohesive $k$-truss than what $u$ belongs to. To achieve this, we introduce the average support of a node into the basic algorithm. Intuitively, if $v$'s every adjacent edge has a large support, then $v$ tends to belong to a $T_k$ with a large $k$. So, we integrate the average support of two nodes $u,v$ with the support of $e_{uv}$ to update the original $p_{uv}$ and optimize the transition matrix $\bm{P}$ in \S \ref{sec:rw_op1}.
	
	
	\vspace{0.1cm}
	\noindent\underline{(3) Optimization with skewness of support (\textbf{\S \ref{sec:rw_op2}}).} We noticed that the above method would face a problem when the support of a node's adjacent edges is extremely skewed, i.e., the average support of a node cannot correctly reflect the community's cohesiveness, as the average support is dominated by those edges with extremely high or low support. To handle this, we leverage the skewness \cite{Doane2011,VonHippel2005,Groeneveld1984} of support of a node's adjacent edges to fine-tune the biased average support. Specifically, we decrease (increase) the average support when a right-skew (left-skew) occurs. We apply this fine-turned average support to optimize $\bm{P}$, thus enhancing the effectiveness.
	
	\vspace{0.1cm}
	\noindent\underline{(4) Optimization with bound of trussness (\textbf{\S \ref{sec:rw_op3}}).} The transition matrix $\bm{P}$ designed above is based on the fundamental concept: support of an edge. Compared with an edge's support, its trussness is the strictest feature to measure a community's cohesiveness. So, it's reasonable to use an edge's trussness to replace its support in the design of $\bm{P}$. However, computing an edge's exact trussness in real-time is impractical for the online random walk algorithm, so we turn to compute an upper bound of an edge's trussness instead, and then we integrate this trussness bound of an edge $e_{uv}$ with the fine-tuned average support of nodes $u,v$ to further optimize $\bm{P}$.
	
	Since the effectiveness of random walk-based algorithms (\S \ref{sec:rw}) depend on the cohesiveness-aware transition matrix $\bm{P}$, we theoretically analyze the rationality of designing transition matrix using the selected cohesiveness features of average support, support skewness, and trussness bound, through \textit{Bayesian theory based on Gaussian Mixture Model with Box-Cox Transformation and Copula Function Fitting} (\textbf{\S \ref{sec:bayesian}}).
	
	Moreover, we extend our random walk-based algorithms twofold. First, we present a lightweight method to iteratively refine the top-$n$ key-members, in an ``expand-replace" manner (\textbf{\S \ref{sec:refinement}}). Experimental study shows that it is quite efficient and effective (e.g., precision approaches to nearly 100\% with 2 iterations of refinement on average). Second, we extend our solution to support CKS with multiple query nodes in \textbf{\S \ref{sec:query_node}}. 
	

	\vspace{0.1cm}
	\noindent\textbf{Contributions.} Our contributions can be concluded as follows.
	\begin{itemize}[leftmargin=*]
		\item We motivate a new problem of community key-members search (CKS) in \textbf{\S \ref{sec:pre_pro}} and present exact algorithms (\textbf{\S \ref{sec:exact_algorithms}-\ref{sec:exact_index}}) atop a TD-based exact framework (\textbf{\S \ref{sec:exact_framework}}).
		\item We propose a basic random walk-based algorithm (\textbf{\S \ref{sec:rw_basic}}) for CKS by considering the cohesiveness feature (i.e., edge's support) in transition matrix design. We next optimize this basic algorithm by updating the transition matrix using more representative cohesiveness features (\textbf{\S \ref{sec:rw_op1}-\ref{sec:rw_op3}}).
		\item We theoretically analyze the rationality of designing transition matrix using the selected cohesiveness features, through Bayesian theory based on Gaussian Mixture Model with Box-Cox Transformation and Copula Function Fitting (\textbf{\S \ref{sec:bayesian}}).
		\item We propose a lightweight refinement method to refine the CKS result with little overhead (\textbf{\S \ref{sec:refinement}}). We extend our solution to support CKS with multiple query nodes (\textbf{\S \ref{sec:query_node}}).
		\item We conduct extensive experiments to evaluate: effectiveness and efficiency (\textbf{\S \ref{sec:effect}-\ref{sec:efficiency}}), case study (\textbf{\S \ref{sec:case}}), parameter sensitivity (\textbf{\S \ref{sec:para}}), and effect of refinement method (\textbf{\S \ref{sec:refine}}), showing our solution's superiority.
	\end{itemize}

	\section{Preliminaries and Problems}
	
	\label{sec:pre_pro}
	
	\subsection{Preliminaries}
	\label{pre}
	
	We consider an undirected, unweighted simple graph $G=(V_G,E_G)$, where $V_G$ ($E_G$) is the node (edge) set. Given a node $u\in V_G$, we denote its neighbors by $N(u)$ and use $deg(u)$ to indicate $u$'s degree, i.e., $deg(u) = |N(u)|$. We use $e_{uv}$ to denote the edge between two nodes $u,v\in V_G$. In the reminder of this paper, we simplify an edge as $e$ unless it's necessary to specify the endpoints $u,v$. We use $\Delta_{uvw}$ to denote the triangle formed by three nodes $u,v,w\in V_G$. Initially, in this paper, we focus on the widely studied homogeneous graphs. In future, we will consider more complex heterogeneous graphs.
	
	\vspace{0.1cm}
	\begin{myDef}
		\label{def:support}
		\textbf{Support} \cite{Yang2020,Wang2012}. Given an edge $e_{uv}\in E_G$, we define $e_{uv}$'s support as the number of triangles containing $e_{uv}$, i.e., $sup(e_{uv})=|\Delta_{uvw}|$, where $w$ is the common neighbor of $u,v$. So, $sup(e_{uv})$ can be computed by $|N(u)\cap N(v)|$.
	\end{myDef}
	
	\vspace{0.1cm}
	\begin{myDef} 
		\label{def:ktruss}
		\textbf{$\bm{k}$-truss} \cite{Yang2020,Wang2012}. Given a graph $G$ and an integer $k\geq 2$, the $k$-truss is defined as the largest subgraph $T_k\subseteq G$ in which each edge's support $sup(e)\geq k-2$.
	\end{myDef}
	
	\vspace{0.1cm}
	\begin{myDef}
		\label{def:trussness}
		\textbf{Trussness} \cite{Wang2012,Huang2014}. Given an edge $e\in E_G$, we define $e$'s trussness $\phi(e)$ as the maximum $k$ of the $k$-truss $T_k$ that $e$ belongs to, i.e., $\phi(e)=\max\{k: e\in E_{T_k}\}$. So, given $ \phi(e)=k$, we have $e\in E_{T_k}$ but $e \notin E_{T_{k'}}$ for any $k'>k$.
	\end{myDef}
	
	\vspace{0.1cm}
	\begin{myExample}
		\label{exp:ktruss}
		Considering the edge $e$ between nodes $u_6$ and $u_8$ in Figure \ref{fig:example}, we have $sup(e)=4$ (i.e., contained by four triangles) and $\phi(e)=5$ because $e$ belongs to a 5-truss formed by $\{u_6,u_7,u_8,u_9,u_{10}\}$, which $k=5$ is the largest.
	\end{myExample}
	
	\subsection{Problem Definition}
	\label{sec:problem}
	Given a $k$-truss $T_k$, some edges in $T_k$ would have trussness $\phi(e)\geq k$. E.g., the 5-truss edges (with trussness as 5) in Figure \ref{fig:example} also belong to a 3-truss. So, we define any edge's maximum trussness in a $T_k$ as $\hat{k}=\max\{\phi(e): e\in E_{T_{k}}\}$, indicating that there exists a $T_{\hat{k}}\subseteq T_k$ having the largest cohesiveness. 
	
	\vspace{0.1cm}
	\begin{myDef}
		\label{def:keymember}
		\textbf{Key-members}. Given a $k$-truss $T_k$, we define the key-members of $T_k$ as the node set $V_{T_{\hat{k}}}$ of the most cohesive $T_{\hat{k}}\subseteq T_k$, where $\hat{k}$ is the maximum trussness of any edge in $T_k$ so that we cannot find a $T_{k'}\subseteq T_k$ with $k'>\hat{k}$. 
	\end{myDef}
	
	\vspace{0.1cm}
	\noindent\textbf{CKS Problem.} Given a graph $G=(V_G,E_G)$ and a query node $q\in V_G$, CKS returns the key-members $V_{T_{\hat{k}}}$ of the $k$-truss $T_k\subseteq G$ that contains $q$, satisfying the following properties:
	\begin{itemize}[leftmargin=*]
		\item \textbf{Participation.} Both the query node $q$ and all key-members belong to the same community $T_k$;
		\item \textbf{Community's maximality.} $T_k$ is the most cohesive $k$-truss that contains $q$ with the maximum $k$, which means we cannot find another $T_{k'}$ containing $q$ with a larger $k'>k$;
		\item \textbf{Key-members' maximality.} The induced graph of key-members, i.e., $T_{\hat{k}}$, is the most cohesive $k$-truss in $T_k$; this means we cannot find another $T_{k'}\subseteq T_k$ with a larger $k'>\hat{k}$.
	\end{itemize}
	
	\section{Exact Baselines}
	\label{sec:exact}
	Before discussing our random walk-based solutions, we introduce an exact framework to solve CKS based on truss decomposition (TD) \cite{Wang2012}. It easily adopts to existing TD methods. For simplicity, we briefly introduce exact algorithms w/o and w/ index based on representative TD methods in \S \ref{sec:exact_algorithms}-\ref{sec:exact_index}. These algorithms server as the exact baselines in our experimental study (\S \ref{sec:exp}), which are simple to implement but costly on efficiency. 
	
	\subsection{An Exact Framework for CKS Problem}
	\label{sec:exact_framework}
	Given a graph $G=(V_G,E_G)$ and a query node $q$, Algorithm \ref{alg:framework} shows the framework consisting of following steps.
	
	\vspace{0.1cm}
	\noindent\textbf{Finding the most cohesive $T_k$.} We find a $T_k\subseteq G$ containing $q$ with the maximum $k$ by TD (line 1). Differing from the classic $k$-truss-based CS problems \cite{Liu2020a,Huang2014,Jiang2021}, we aim to find $T_k$ with the largest $k$, rather than a user-specific $k$.
	
	\vspace{0.1cm}
	\noindent\textbf{Finding key-members from $T_k$.} We next take $T_k$ as input to find the most cohesive $k$-truss $T_{\hat{k}}\subseteq T_k$ with the maximum $\hat{k}$ by the same TD method used in the first step (line 2). Then, we return the node set of $T_{\hat{k}}$ as the key-members (line 3). 
	
	\subsection{Truss Decomposition-based Exact Algorithm}
	\label{sec:exact_algorithms}
	We easily can provide an exact algorithm by simply deploying any existing TD method on the exact framework. In experimental study, we establish three exact algorithms based on TD-bottomup \cite{Wang2012}, TD-topdown \cite{Wang2012}, and AccTD \cite{Che2020}, denoted by Exact-TD-bottomup, Exact-TD-topdown, and Exact-AccTD for evaluation. Since TD-bottomup and TD-topdown are the fundamental of AccTD, we take them as an examples to show their procedures and complexities. 
	

	\setlength{\textfloatsep}{0cm}
	\setlength{\floatsep}{0.1cm}
	\setlength{\intextsep}{0cm}
	\begin{algorithm} [t]
		\small
		\setstretch{0.95}
		\SetAlgoNoLine
		\LinesNumbered
		\caption{An exact framework for CKS} 
		\label{alg:framework}
		\KwIn{A simple graph $G = (V_G,E_G)$, a query node $q$} 
		\KwOut{Key-members $V_{T_{\hat{k}}}$} 
		$T_k \leftarrow $ use TD to find a $T_k\subseteq G$ containing $q$ with the maximum $k$\;
		$T_{\hat{k}} \leftarrow $ use TD to find a $T_{\hat{k}}\subseteq T_k$ with the maximum $\hat{k}$\;
		\Return{$V_{T_{\hat{k}}}$}
	\end{algorithm}
	
	\vspace{0.1cm}
	\noindent\textbf{Exact-TD-bottomup.} Given a graph $G$ and a query node $q$, Exact-TD-bottomup calls TD-bottomup to find the most cohesive $T_k\subseteq G$ (step 1). Specifically, TD-bottomup initializes the support of each edge in $G$. Then, it starts from the smallest $k$, i.e., $k=3$, to iteratively remove all edges with support $sup(e)< k-2$ from $G$. Removing an edge $e_{uv}$ may invalidate all triangles consisting of $e_{uv}$, i.e., $\Delta_{uvw}$, where $w$ is one common neighbor of $u,v$. So, we need to decrease the support of the other two edges $e_{uw}$ and $e_{vw}$ for each $\Delta_{uvw}$, and iteratively check whether they would be removed from $G$. This process continues until all edges with support $sup(e)<k-2$ are removed and the remaining edges form a $k$-truss. TD-bottomup stops when the found a $k$-truss does not contain $q$. It means that the $(k-1)$-truss found in the last iteration is the most cohesive $T_k$ with the largest $k$ that contains $q$. Otherwise, TD-bottomup processes the next iteration of $k+1$. In step 2, Exact-TD-bottomup calls TD-bottomup to find the most cohesive $T_{\hat{k}}\subseteq T_k$ from the $k$ of $T_k$. Here, we do not need to check if each found $k$-truss contains $q$. TD-bottomup terminates when all edges are removed at a certain iteration of $k$. It means that we cannot find a $k$-truss at this iteration so that the previous found $(k-1)$-truss is the most cohesive $T_{\hat{k}}$. Otherwise, we move to process the next iteration of $k+1$.
	
	\vspace{0.1cm}
	\noindent\textbf{Complexity.} Since Exact-TD-bottomup enumerates all subgraphs containing $q$ and deletes edges to find a $k$-truss from the smallest $k=3$, its time complexity is $O(\sum_{H\subseteq G}|E_{H}|^{1.5})$, where $H\subseteq G$ is a subgraph of $G$. For each $H$, it requires up to $O(|E_{H}|^{1.5})$ time for computing a $k$-truss \cite{Wang2012,Liu2020}. Next, we repeat it to find $k$-truss $T_{\hat{k}}\subseteq T_k$ from the $k$ of $T_k$, so it costs $O(\sum_{H\subseteq T_k}|E_{H}|^{1.5})$. So, the total complexity is $O(\sum_{H\subseteq G}|E_{H}|^{1.5}+\sum_{H\subseteq T_k}|E_{H}|^{1.5})$. In the worst case, we need to enumerate $2^{|V_G|-1}$ subgraphs (that contain $q$) in step 1 and $2^{|V_{T_k}|}$ subgraphs (that do not have to contain $q$) in step 2. Since $T_k|$ is usually significantly smaller than $G$, the first step is more efficient than the second step in practice.
	
\vspace{0.1cm}
\noindent\textbf{Exact algorithm with TD-topdown.} Exact-TD-topdown is established on the basis of the classic top-down TD method \cite{Wang2012}. First, Exact-TD-topdown calls TD-topdown to find the most cohesive $T_k\subseteq G$ (step 1). It still needs to initialize each edge's support in $G$. Then, it starts from an upper bound of $k$ to check if there exists a $k$-truss containing query node $q$. The upper bound of $k$ can be simply configured as $k=\max\{sup(e_{qu})|u\in N(q)\}+2$, where $e_{qu}$ is an adjacent edge of $q$ and $N(q)$ indicates $q$'s neighbors. In each iteration of $k$, TD-topdown extracts all the triangles that contain those edges with $sup(e)=k$ to form a temporary subgraph. Next, TD-topdown iteratively removes all edges with support $sup(e)<k-2$ from this subgraph. Similar to TD-bottomup, we need to iteratively check whether to remove the other triangle edges that contains the removed $e$ should be removed. This process continues until all edges with support $sup(e)<k-2$ are removed and the remaining edges can form a $k$-truss. If this $k$-truss contains $q$, then it is the desired $T_k$. Otherwise, TD-topdown processes the next iteration of $k-1$. In the second step, Exact-TD-topdown calls TD-topdown to find the most cohesive community $T_{\hat{k}}\subseteq T_k$ by starting with the $k=\max\{sup(e)|e\in E_{T_{k}}\}+2$. TD-topdown terminates when a $k$-truss is found at the iteration of $k$. Otherwise, it moves to process the next iteration of $k-1$.

\vspace{0.1cm}
\noindent\textbf{Complexity.} According to the analysis of \cite{Wang2012}, TD-bottomup and TD-topdown show the same time complexity on TD. So, we take TD-bottomup as an example to show the overall complexity of Exact-TD-bottomup. In the first step, it enumerates all subgraphs containing $q$ and deletes edges to find a $k$-truss from the smallest $k=3$, until the largest $k$ is reached. Hence, the time complexity is $O(\sum_{H\subseteq G}|E_{H}|^{1.5})$, where $H\subseteq G$ is a subgraph of $G$. For each $H$, it requires up to $O(|E_{H}|^{1.5})$ time for computing a $k$-truss \cite{Wang2012,Liu2020}. While in the second step, we repeat this operation on the found $k$-truss $T_k\subseteq G$ from the $k$ of $T_k$, so it costs $O(\sum_{H\subseteq T_k}|E_{H}|^{1.5})$. Thus, the complexity of Exact-TD-bottomup is $O(\sum_{H\subseteq G}|E_{H}|^{1.5}+\sum_{H\subseteq T_k}|E_{H}|^{1.5})$. In the worst case, we need to enumerate $2^{|V_G|-1}$ subgraphs (that must contain $q$) in the first step and $2^{|V_{T_k}|}$ subgraphs (that do not have to contain $q$) in the second step. Since $|V_{T_k}|$ is usually significantly smaller than $|V_G|$, the second step is much more efficient than the first step in practice.


\vspace{0.1cm}
\noindent\textbf{Remarks.} AccTD \cite{Che2020} is the state-of-the-art work that focus on improving TD's efficiency on large-scale dataset. As claimed in \cite{Che2020}, AccTD has the same time complexity bound as TD-bottomup and TD-topdown. However, AccTD's optimization strategies significantly reduce the practical workload. In \S \ref{sec:exp}, we also implement an exact algorithm Exact-AccTD for experimental evaluation.	
	
	\subsection{Exact Algorithm with Index}
	\label{sec:exact_index}
	In the literature of CS, trussness-based index is often used to improve CS's efficiency \cite{Liu2020a,Huang2014,Jiang2021}. For example, the \textbf{T}riangle \textbf{C}onnectivity \textbf{P}reserved Index (TCP-Index) \cite{Jiang2021,Huang2014} is one representative trussness-based index. We can quickly find a certain $k$-truss that contains a query node $q$ via TCP-Index. Thus, we have another exact algorithm with TCP-Index, denoted by Exact-TCP-Index. We only provide a high-level idea of the index construction and refer interested readers to \cite{Huang2014,Jiang2021} for more details. Actually, the TCP-Index for $G$ is a set of TCP-Indices for all nodes of $V_G$. For each node $u\in V_G$, we first extract all neighbors of $u$ as $N(u)$ to form the induced graph of $N(u)$, denoted by $G_u$. Next, we assign a weight on each edge in $G_u$ by this edge's trussness. Third, we generate a maximum spanning forest of this weighted $G_u$ as the TCP-Index for node $u$, denoted by $\mathcal{T}_u$. We repeat this for every node $u\in V_G$ and return the combination of all nodes' TCP-Indices as the TCP-Index for $G$, denoted by $\mathcal{T}_G=\{\mathcal{T}_u|u\in V_G\}$.
	
	\vspace{0.1cm}
	\noindent\textbf{Exact-TCP-Index.} First, we use the offline built TCP-Index $\mathcal{T}_G$ to return the most cohesive $T_k$ with the largest $k$ that contains $q$. Specifically, we search the TCP-Index of $q$, i.e., $\mathcal{T}_q\in \mathcal{T}_G$ as follows. (1) We select a neighbor $u$ of $q$ as an entry node for searching $\mathcal{T}_q$, satisfying that $e_{qu}$ has the largest trussness. (2) We traverse $\mathcal{T}_q$ from $u$ to collect all nodes connected by edges with trussness $\phi(e_{qu})$. Then, we repeat above operations by continuously searching on these nodes' TCP-Indices until all nodes connected by edges with trussness $\phi(e_{qu})$ are found. As a result, all found nodes and $q$ form the most cohesive $T_k$ (with the largest $k=\phi(e_{qu})$) that we desire. Next, we use TCP-Index to return the most cohesive $k$-truss $T_{\hat{k}}\subseteq T_k$. We first enumerate all edges in $T_k$ to find an edge $e_{uv}$ with the largest trussness, which is exactly the maximum $\hat{k}$ of $T_{\hat{k}}$ that we desire. Finally, we take an arbitrary endpoint of $e_{uv}$ and its TCP-Index as input and repeat the same procedure of the first step to find all the nodes connected by edges with trussness $\hat{k}$. As a result, all found nodes and the selected endpoint are key-members that we are looking for.
	
	\vspace{0.1cm}
	\noindent\textbf{Complexity.} According to \cite{Huang2014}, the TCP-Index for a graph $G$ can be constructed in $O(\sum_{e_{uv}\in E_{G}}\min\{deg(u),deg(v)\})$ time and $O(|E_G|)$ space. The searching time of Exact-TCP-Index is dominated by the the size of TCP-Index for $G$. In the worst case, we require $O(|E_{\mathcal{T}_G}|)$ time to access $\mathcal{T}_G$ to find $T_k$, and we need additional $O(|E_{\mathcal{T}_G}|)$ time to find $T_{\hat{k}}$. This is because $T_{\hat{k}}\subseteq T_k$ with a smaller size, so the search time of second step is also bounded by $O(|E_{\mathcal{T}_G}|)$.
	
	\section{Random Walk-based Algorithms}
	\label{sec:rw}
    Exact algorithms proposed in \S \ref{sec:exact} are simple to implement but costly on efficiency or introduce additional overhead for index storage. This motivates us to present random walk-based algorithms with several optimizations to achieve a good balance between effectiveness and efficiency in \S \ref{sec:rw_basic}-\ref{sec:rw_op3}. 
	
	
	\subsection{Basic Algorithm}
	\label{sec:rw_basic}
	Random walk is popular for graph sampling as its scalability and simplicity of implementation \cite{Zhao2015,Li2019,Wang2022a}. A general random walk on a graph $G$ can be modeled as a finite Markov Chain \cite{Li2019}. A walker starts from a node $u_0\in V_G$, chooses a neighbor of $u_0$ and moves to it with the transition probability defined in the transition matrix $\bm{P}=|V_G|\times |V_G|$. It continues to walk until a stationary distribution $\bm{\pi}=\{\pi_1,\cdots,\pi_{|V_G|}\}$ is reached, where $\sum \pi_i=1$ and $\pi_i$ is the stationary visiting probability of each $u_i\in V_G$ when random walk converges. Recall the CKS's definition, we aim to find key-members belonging to the most cohesive $T_{\hat{k}}$ within a query node $q$'s cohesive community $T_k$. Thereby, it is reasonable to design a $\bm{P}$ based on some representative cohesiveness features, so that the random walk would converges to a stationary distribution of which key-members may have larger visiting probabilities than others. We first present an original version of our basic algorithm, then provide a relaxed version that is more efficient in practice.
	
	\vspace{0.1cm}
	\noindent\textbf{Original version.} Given a graph $G$ and a query node $q$, our basic algorithm has four steps: (1) extract an $m$-bounded subgraph $G_q$ of $q$ from $G$, (2) design $\bm{P}$ over $G_q$, (3) random walk until it converges, and (4) return top-$n$ key-members. 
	
	\vspace{0.1cm}
	\noindent\underline{(1) \textit{Extract am $m$-bounded subgraph.}} According to small world theory \cite{Kleinberg2000,Malkov2020}, two nodes in the same community exhibit strong access locality \cite{Yang2012}, which means two nodes are more likely to belong to the same community if they are located in each other's localized space. So, we assume that key-members can be found in an $m$-bounded subgraph of the query node $q$, denoted by $G_q\subseteq G$, and we conduct the random walk over $G_q$ instead of the entire $G$. All nodes in $G_q$ are within $m$-hops from $q$, which can be found quickly through a BFS starting from $q$. In this BFS, we must ensure that for each visited node $v$, it has at least one common neighbor with its parent node $u$, i.e., $sup(e_{uv})\geq 1$. This is important for the random walk to converge (Lemma \ref{lemma:irreducible}). We will discuss in Remarks part why exact algorithm (\S \ref{sec:exact_algorithms}) cannot be benefited from the $m$-bounded subgraph, mainly because it would diminish their effectiveness significantly.
	
	\vspace{0.1cm}
	\noindent\underline{(2) \textit{Design transition matrix.}} According to Definition \ref{def:ktruss}, each edge $e$ in a $T_k$ must have a support $sup(e)\geq k-2$, so $k-2$ is a lower bound of support for each edge in $T_k$. A larger $k$ indicates that each edge's lower bound of support is larger than that of a smaller $k$. Thus, a simple idea is to use each edge's support as the cohesiveness feature to roughly measure whether it belongs to a $k$-truss with a larger $k$ or a smaller $k$. Given two edges $e, e'$ from different $k$-truss $T_{k}$ and $T_{k'}$, it's reasonable to say that $sup(e)\geq sup(e')$ holds with a relatively higher probability, if $k\geq k'$. Following this assumption, in a random walk, if a walker towards along a path consisting of edges with large support as much as possible, then this walker is more likely to reach to a $k$-truss with a large $k$. So, as the first step, we design a transition matrix $\bm{P}=[p_{ij}]$ based on edges' support (Eq. \ref{eq:transition1}), where $p_{ij}$ is the transition probability of moving from node $u_i$ to $u_j$, $N(u_i)$ is the neighbors of $u_i$, and $sup(e_{ix})$ is the support of edge $e_{ix}$ between $u_i$ and its neighbor $u_x\in N(u_i)$. In this way, we prefer to choose an edge with a larger support to move at each walk step. 
	
	\begin{equation}
		\label{eq:transition1}
		p_{ij} = \frac{sup(e_{ij})}{\sum_{u_x \in N(u_i)} sup(e_{ix})}
	\end{equation}
	
	\begin{myExample}
		\label{exp:rw_basic}
		In Figure \ref{fig:example}, edges between $u_{14}$ and its neighbors $\{u_{9},u_{10},u_{13}\}$ have a support of 2, while other adjacent edges have a support of 1. So, the transition probability from $u_{14}$ to $\{u_{9},u_{10},u_{13}\}$ is $\frac{2}{2*3+1*2}=25\%$ and that of other edges is $12.5\%$. Thus, a walker at node $u_{14}$ has $50\%$ total probability of moving to the most cohesive 5-truss via visiting $\{u_9,u_{10}\}$.
	\end{myExample}
	
	A random walk can converge to a stationary distribution only if the finite Markov Chain (MC) is \textit{irreducible} and \textit{aperiodic} \cite{Ross2014}. We next show our random walk can converge in the following two Lemmas.
	
	\vspace{0.1cm}
	\begin{myLemma}
		\label{lemma:irreducible}
		Our semantic-aware random walk is irreducible.
	\end{myLemma}
	
	\vspace{0.1cm}
\begin{IEEEproof}
\label{pf:irreducible}
An MC is irreducible if any two nodes are reachable in finite steps. So, this Lemma naturally holds because each edge has a non-zero transition probability. 
\end{IEEEproof}
	
	In an MC, each node has period $k$ if any return to itself must occur in multiples of $k$ steps, and an MC is aperiodic if it has at least one node having period one \cite{Ross2014}. To satisfy this, we follows \cite{Wang2022a} to change $G_q$ with a small modification: We add a self-loop edge on the query node $q$ with a small fake transition probability $p_{qq}$ (e.g., 0.001). A walker starting from $q$ tends to walk outward rather than be stuck at $q$ due to this small $p_{qq}$, and it has little effect on the convergence time. It is easy to verify that our random walk is aperiodic. 
	
	\vspace{0.1cm}
	\begin{myLemma}
		\label{lemma:aperiodic}
		Our semantic-aware random walk is aperiodic.
	\end{myLemma}
	
	\vspace{0.1cm}
	This directly holds as the self-loop edge has period one. 
	
	\vspace{0.1cm}
	\noindent\underline{(3) \textit{Random walk until convergence.}} Given a transition matrix $\bm{P}$ over an $m$-bounded subgraph $G_q$, and a query node $q$, we use matrix multiplication to update the stationary distribution as follows. First, we initialize the stationary distribution $\bm{\pi}=\{\pi_q,\pi_1,\dots,\pi_{|V_{G_q}|-1}\}$ at the first iteration as $\bm{\pi}^{(0)}=\{1,0,\dots,0\}$, where $\pi_q=1$ because we start the random walk from $q$. Second, we apply Eq. \ref{eq:matrix_product} to update $\bm{\pi}$ at the $t$-th iteration, denoted by $\bm{\pi}^{(t)}$, based on the $\bm{\pi}^{(t-1)}$ obtained at the $(t-1)$-th iteration. The random walk converges when $\bm{\pi}$ is no longer changing, i.e., $\bm{\pi}^{(t)}=\bm{\pi}^{(t-1)}$. 
	\begin{equation}
		\label{eq:matrix_product}
		\bm{\pi}^{(t)}=\bm{\pi}^{(t-1)}\times \bm{P}
	\end{equation}
	
	\noindent\underline{(4) \textit{Return top-$n$ key-members.}} We obtain a stationary distribution $\bm{\pi}$ after random walk converges. Since we design $\bm{P}$ based on the cohesiveness feature, i.e., edge support, it is more likely that a node $u_i$ from a $T_k$ with a large $k$ would be visited with a large stationary visiting probability of $\pi_i$. So, we return the top-$n$ nodes with greater $\pi_i$ as key-members. In \S \ref{sec:exp}, we show the effect of $n$ on the effectiveness.
	
	\begin{figure}
	\vspace{-0.3cm}
		\setlength{\abovecaptionskip}{0.1cm}
		\centering
		\subfigure[Efficiency w.r.t. iterations]{
			\includegraphics[scale=0.13]{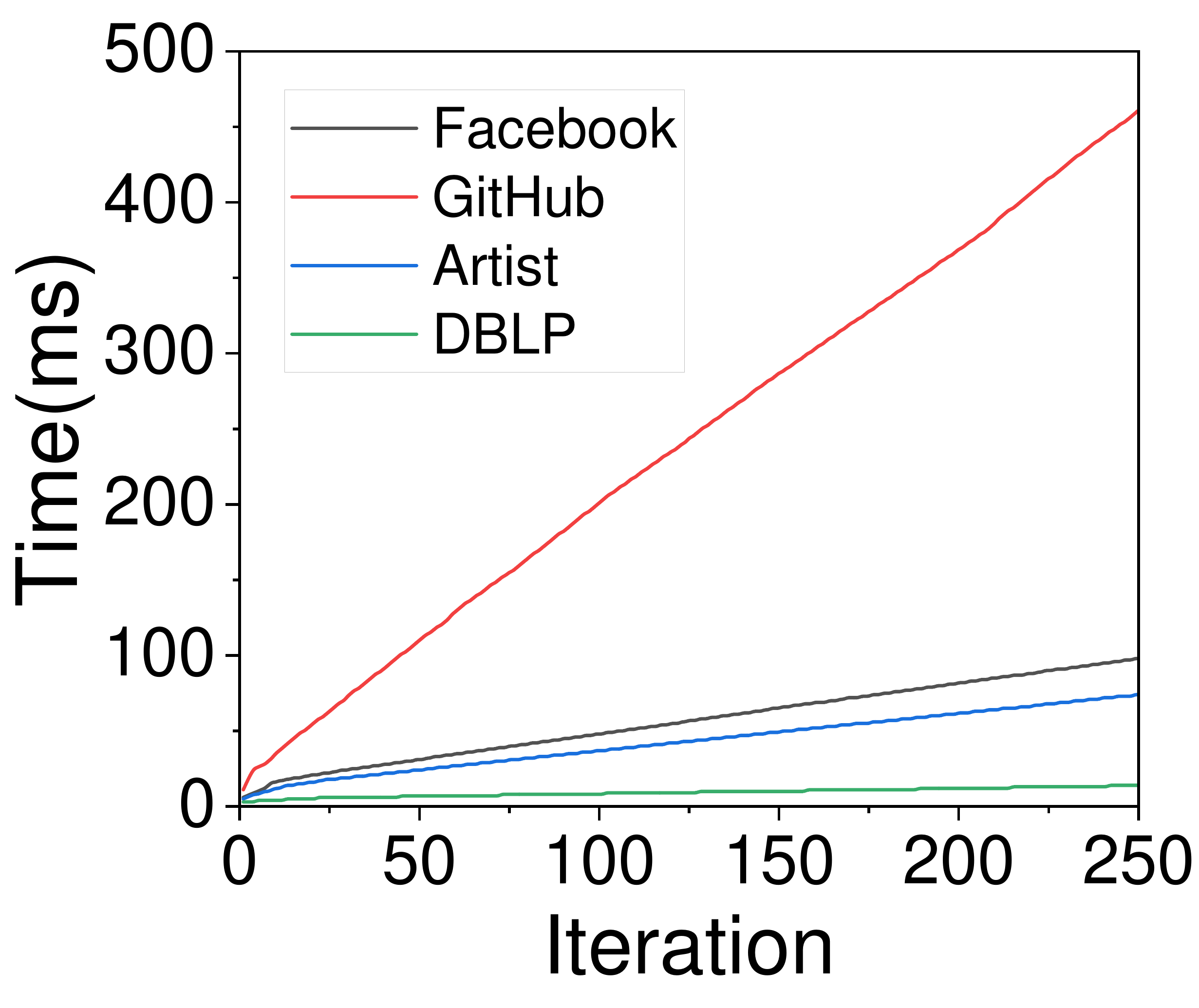}
			\label{fig:rw_effect1}
		}%
		\subfigure[Effectiveness w.r.t. iterations]{
			\includegraphics[scale=0.13]{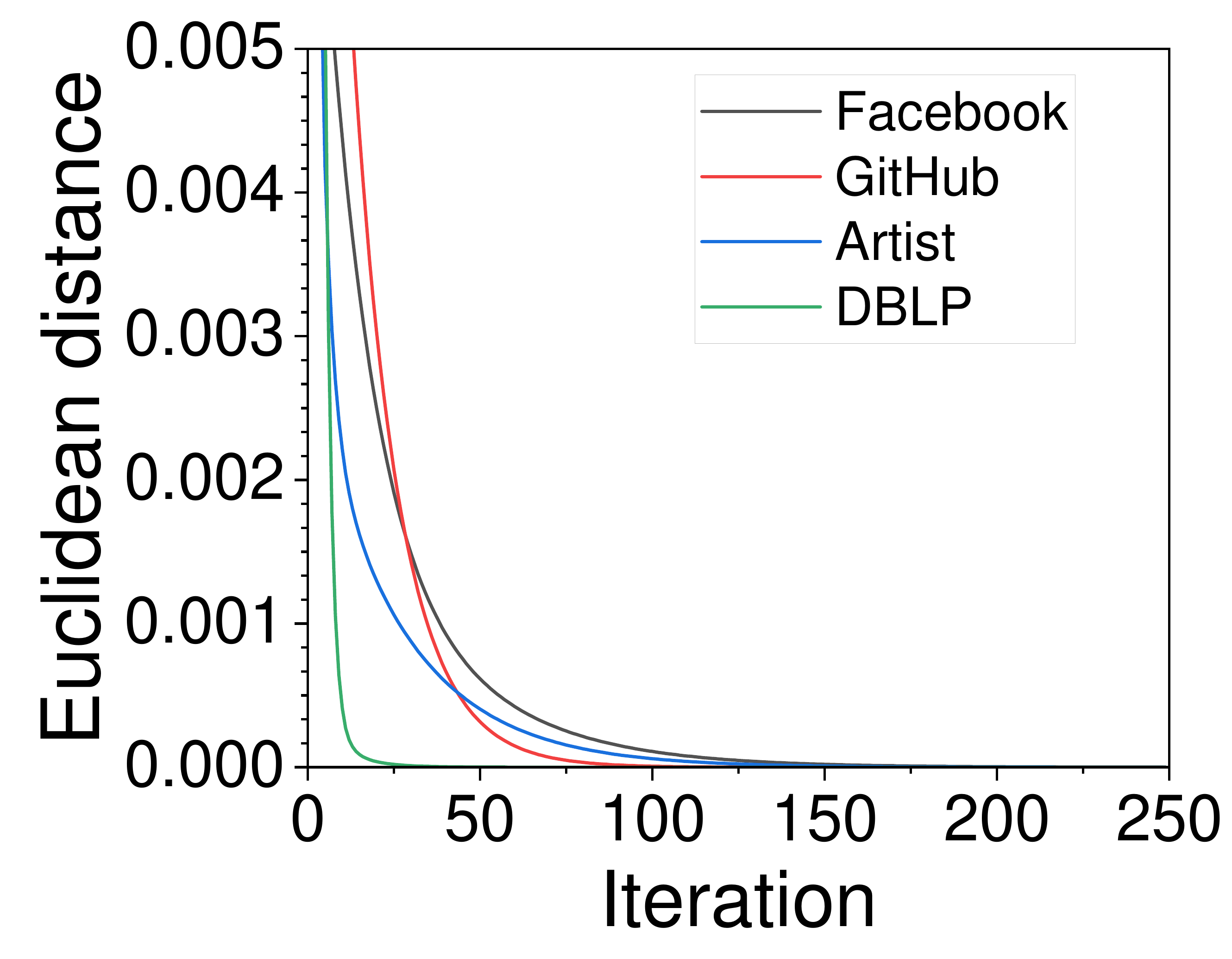}
			\label{fig:rw_effect2}
		}%
		\caption{Effect of iterations on the basic random walk}
		\label{fig:rw_effect}
	\end{figure}
	
	\vspace{0.1cm}
	\noindent\textbf{Relaxed version.} Since we apply matrix multiplication to iteratively update $\bm{\pi}$ until it converges, the more iterations, the more time is required. Figure \ref{fig:rw_effect}(a) shows the effect of iterations on random walk's efficiency. For each dataset, we process the original version algorithm 100 times with randomly selected query nodes. The converge time increases as iteration increases. We also provide the average Euclidean distance between $\bm{\pi}^{(t)}$ and $\bm{\pi}^{(t-1)}$ in Figure \ref{fig:rw_effect}(b), from where we find that the distance decreases as iteration increases, and tends to be stable after 150 iterations. This inspired us to use an approximate stationary distribution instead of the exact stationary distribution, as they have minor difference and offer a good efficiency. So, in Algorithm \ref{alg:relax_rw}, we relaxed the original version by changing the terminate condition from the complete convergence to a fixed \# iterations $r$ is reached (lines 7-8). In \S \ref{sec:exp}, we show the effect of $r$ on CKS's performance. 
	
	\begin{algorithm} [t]
		\setstretch{0.95}
		\small
		\LinesNumbered
		\caption{{\small Relaxed basic random walk-based algorithm}}
		\label{alg:relax_rw}
		\KwIn{$G=(V_G,E_G), q, m, n, r$} 
		\KwOut{top-$n$ key-members}
		$t=1$, $\bm{\pi}^{(t-1)}=\{1,0,\dots,0\}$\;
		\tcp{extract $m$-bounded subgraph}
		$G_q=$ getBoundedGraph($G,q,m$)\;
		\tcp{transition matrix initialization}
		\For{$\forall e_{ij}\in E_{G_q}$} 
		{ 
			$sup(e_{ij}) = |N(u_i) \cap N(u_j)|$\;
		}
		\For{$\forall e_{ij}\in E_{G_q}$}
		{
			$p_{ij}=\frac{sup(e_{ij})}{\sum_{u_x \in N(u_i)} sup(e_{ix})}$; \tcp*[h]{Eq. \ref{eq:transition1}} \\
		}
		\tcp{random walk}
		\While{$t++\leq r$} 
		{ 
			$\bm{\pi}^{(t)}=\bm{\pi}^{(t-1)}\times \bm{P}$; \tcp*[h]{Eq. \ref{eq:matrix_product}} \\
		} 
		\Return top-$n$ nodes from $\bm{\pi}$ with the greatest $\pi_i$\;
	\end{algorithm}
	
	\vspace{0.1cm}
	\noindent\textbf{Complexity.} The total time of relaxed version is $O(|E_{G_q}|+|V_{G_q}|+|E_{G_q}|+r|V_{G_q}|)$, where $|E_{G_q}|$ and $|V_{G_q}|$ are \# edges and \# nodes in the $m$-bounded subgraph $G_q$. We need $O(|E_{G_q}|+|V_{G_q}|)$ time to extract $G_q$. We then initialize $\bm{P}$ by computing the transition probabilities for all $|E_{G_q}|$ edges. The time of matrix multiplication for $r$ iterations is $r|V_{G_q}|$.
	
\begin{figure}
		\setlength{\abovecaptionskip}{0.1cm}
		\centering
		\includegraphics[scale=0.33]{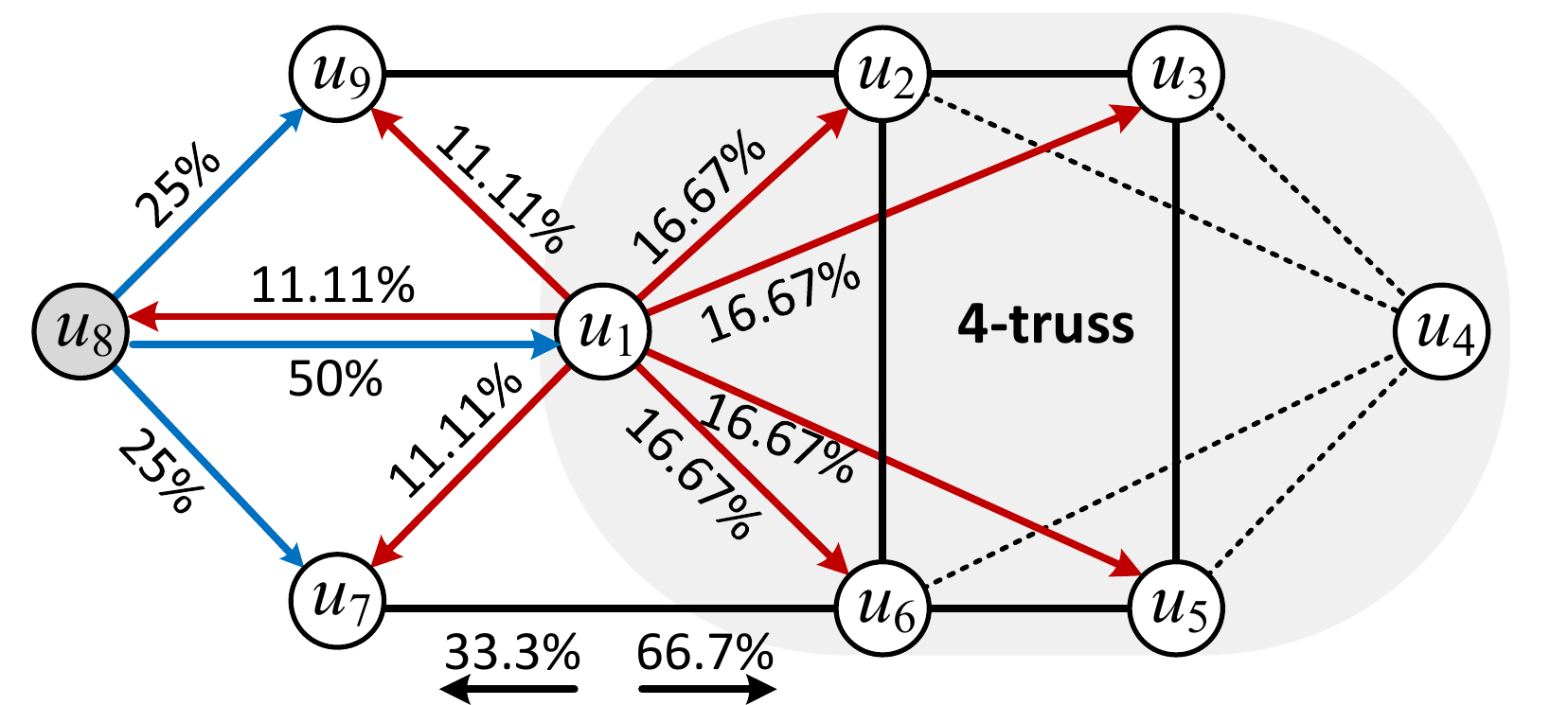}
		\caption{The effect of $m$-bounded subgraph}
		\label{fig:m_graph}
	\end{figure}

	\vspace{0.1cm}
	\noindent\textbf{Remarks.} The $m$-bounded subgraph is very helpful to improve efficiency, as it reduces the walk space from a large $G$ to a small $G_q$. However, it cannot be adopted in the exact algorithm because it would greatly undermine the effectiveness. We explain this by an example in Figure \ref{fig:m_graph}. Given $q=u_8$, we find key-members as $\{u_1,u_2,u_3,u_4,u_5,u_6\}$, as they participate in the most cohesive 4-truss within the community containing $u_8$. If we apply the $m$-bounded subgraph in the exact algorithm (e.g., $m=2$), then $u_4$ is excluded from $G_q$ and we will obtain the key-members as $\{u_1,u_2,u_3,u_5,u_6,u_7,u_8,u_9\}$, which is quite different from the original result. Back to our basic solution, first we have $50\%$ probability to move to $u_1$ from $u_8$, then we have $66.7\%$ probability in total to move to the 4-truss from $u_1$, which is larger than the total probability ($33.3\%$) of coming back to $\{u_7$,$u_8$,$u_9\}$. Finally, we find $\{u_1$,$u_2$,$u_3$,$u_5$,$u_6\}$ as key-members according to their higher stationary visiting probabilities than others. we show the effect of $m$ in \S \ref{sec:para}.
	
	\vspace{-0.1cm}
	\subsection{Optimization with Average Support}
	\label{sec:rw_op1}
	The basic algorithm performs well in many scenarios, except the case where one node has a large number of adjacent edges but most of them have small supports. Given the graph shown in Figure \ref{fig:average_support}, $u_1,u_2$ have 8 common neighbors, so the edge $e_{12}$ has a support $sup(e_{12})=8$. However, $u_2$ only belongs to a $3$-truss $T_3$, because other adjacent edges of $u_2$ only have $sup(\cdot)=1$. If we apply the basic algorithm on this graph, then $u_2$ would have a large stationary visiting probability, because the transition probability $p_{12}=\frac{8}{36}=22.22\%$ (Eq. \ref{eq:transition1}) is higher than others, making the random walk tends to back to $u_2$ than walking outward to the right part. Ideally, if a node's every adjacent edge has a large support, then it tends to belong to a $T_k$ with a large $k$. So, we define a node's average support by considering this node's global support information and use it as a complement to edge support to optimize $\bm{P}$.
	
	\vspace{0.1cm}
	\begin{myDef} 
		\label{def:average_support}
		\textbf{Average Support}. Given a node $u_i\in V_G$, we define $u_i$'s average support as $A(u_i)=\frac{\sum_{u_x\in N(u_i)}sup(e_{ix})}{|N(u_i)|}$.
	\end{myDef}
	
	\vspace{0.1cm}
	Given two nodes $u_i,u_j$ with average support $A(u_i)$ and $A(u_j)$, the transition probability $p_{ij}$ is computed by Eq. \ref{eq:transition2}. Here, we consider both the edge support $sup(e_{ij})$ and average support of $u_i,u_j$, to represent a community's cohesiveness. It tends to walk from $u_i$ to $u_j$ when $sup(e_{ij})$ and $\frac{A(u_j)}{A(u_i)}>1$ are large. The larger the $\frac{A(u_j)}{A(u_i)}$, the higher likelihood that $u_j$ belongs to a more cohesive community than that of $u_i$.
	\begin{equation}
		\label{eq:transition2}
		p_{ij} = \frac{sup(e_{ij})\cdot \frac{A(u_j)}{A(u_i)}}{\sum_{u_x \in N(u_i)} sup(e_{ix})\cdot \frac{A(u_x)}{A(u_i)}}
	\end{equation}
	
	\begin{myExample}
		\label{exp:average_support}
		Figure \ref{fig:average_support} shows the transition probabilities computed by Eq. \ref{eq:transition2}. The average support of several nodes are provided in Figure \ref{fig:average_support}, e.g., $A(u_1)=\frac{5*4+8*1+1*8}{14}=2.57$. Based on this information, we compute each edge's transition probability, e.g., $p_{12}$ = $\frac{8\cdot 1.78/2.57}{8 (1\cdot \frac{1}{2.57})+8\cdot \frac{1.78}{2.57}+5 (4\cdot \frac{4}{2.57})}$ = $13.91\%$ and $p_{13}$ = $\frac{4\cdot 4/2.57}{8 (1\cdot \frac{1}{2.57})+8\cdot \frac{1.78}{2.57}+5 (4\cdot \frac{4}{2.57})}$ = $15.65\%$. It has $13.91\%$ probability of moving from $u_1$ to $u_2$ which is smaller than that of moving to $u_3$ ($15.65\%$). From a macro point of view, it is more likely to head toward to the $6$-truss via edges $e_{13}-e_{17}$ with a total probability of $15.65\%\times 5$ = $78.25\%$ and only has a probability of $21.75\%$ to walk toward to the left part.
	\end{myExample}
	
	\subsection{Optimization with Skewness of Support}
	\label{sec:rw_op2}
\begin{figure}
		\setlength{\abovecaptionskip}{0.1cm}
		\centering
		\includegraphics[scale=0.3]{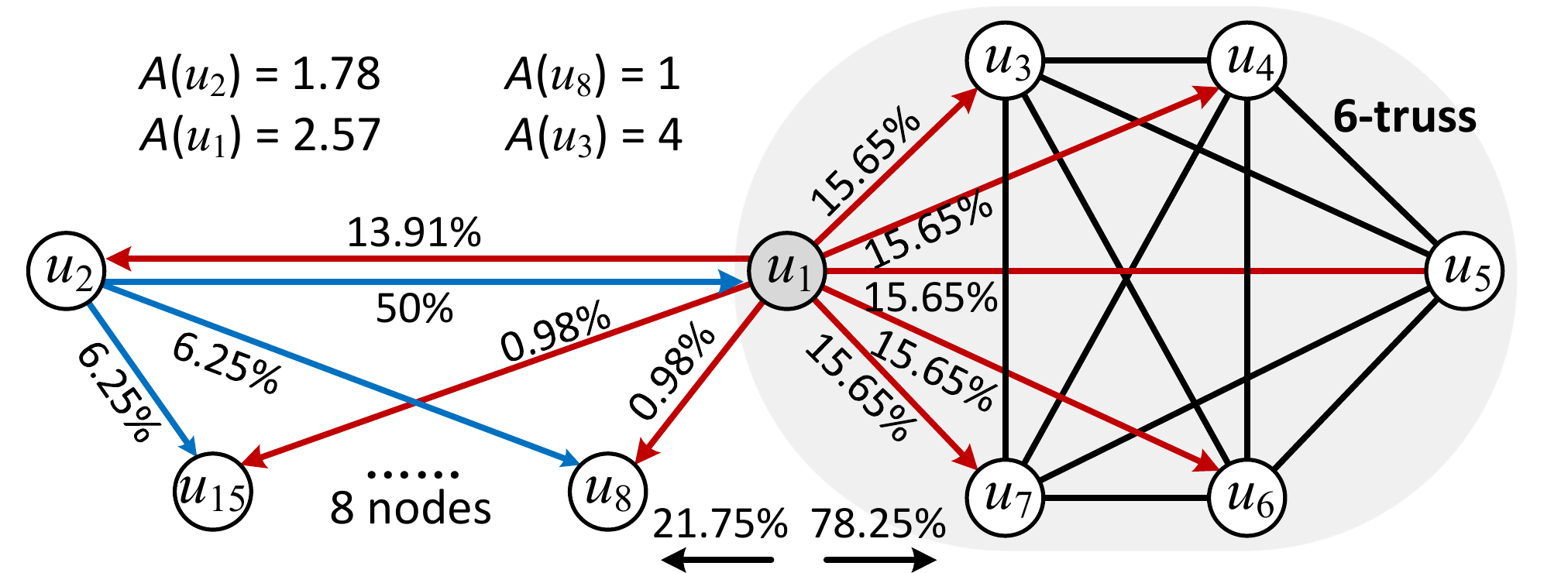}
		\caption{An example of optimization with average support}
		\label{fig:average_support}
	\end{figure}

	The above method would encounter a problem when there is an extremely skewed difference in the support of a node's adjacent edges. In this case, a node's average support inaccurately reflects its community's cohesiveness, as the average support is dominated by those edges with extremely high or low support. This  would erroneously guide a walker toward a less cohesive community by considering an inflated average support, or avoid walking to a more cohesive community by considering an deflated average support. In real-world datasets, we found many of such skewed nodes. Figure \ref{fig:average_support} shows the support distribution of a node with ID 27803 in \texttt{GitHub} dataset, the most cohesive community it belongs to is a 13-truss, but it has an inflated average support of 22 as its average support is dominated by 3\% of edges with support $>$ 100.
	
\begin{figure}
		\setlength{\abovecaptionskip}{-0.2cm}
		\vspace{-0.6cm}
		\centering
		\includegraphics[scale=0.18]{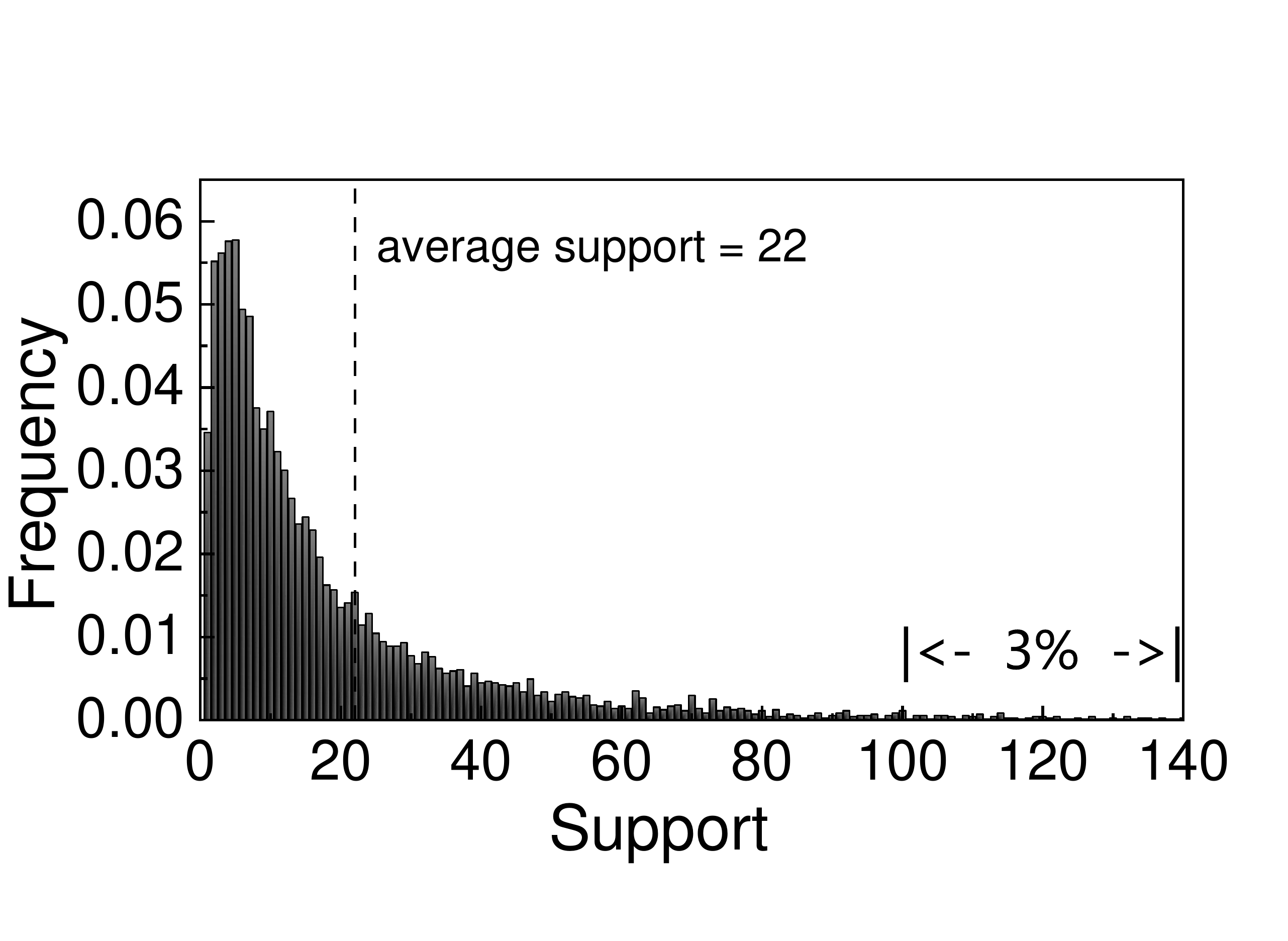}
		\caption{Support distribution of a node in GitHub (ID 27803)}
		\label{fig:support_distribution}
	\end{figure}

	Our solution is to fine-tune the average support based on the support skewness of a node's adjacent edges. Generally, skewness is a measure of the asymmetry of a distribution \cite{Doane2011,VonHippel2005,Groeneveld1984}. A distribution is asymmetrical when its left and right sides are not mirror images. For a right (left)-skewed distribution, it has a longer tail on the right (left) side of its peak \cite{skewness}. If a node's adjacent edges' support follows a right-skewed distribution, then the average support is being skewed to the right of the data center \cite{Dean2018}. So, those edges with extremely large supports (values in the tail) affect the average support more than others, leading to an inflated average support. We should decrease (increase) the average support when right (left)-skew occurs. Figure \ref{fig:support_distribution} shows an example of typical right-skewed distribution where the average support is inflated by a small number of edges with extremely large support, indicating that we should decrease it to some extent. Given a node $u$, in this paper we compute the support skewness for $u$, denoted by $skew(u)$, through \textit{Fisher's moment coefficient of skewness} \cite{Doane2011,Groeneveld1984} by Eq. \ref{eq:skewness}, where $sup(e)$ is the support of $u$'s one adjacent edge and $\sigma$ is the standard deviation of the support of $u$'s adjacent edges. 
	\begin{equation}
		\label{eq:skewness}
		skew(u)=E\left[\left(\frac{sup(e)-A(u)}{\sigma}\right)^3\right]
	\end{equation}
	
	We say a node $u$'s support distribution is right (left)-skewed if $skew(u)>0$ ($skew(u)<0$). For nodes with $skew(u)>0$ ($skew(u)<0$), we decrease (increase) $A(u)$ by Eq. \ref{eq:update_skewness}. The large the $|skew(u)|$, the more the decrement or increment in $A(u)$. If $skew(u)$ = $0$, we keep the average support unchanged.
	\begin{equation}
		\label{eq:update_skewness}
		A^*(u)=A(u)\cdot (1+\alpha\cdot (\frac{1}{1+e^{skew(u)}}-\frac{1}{2}))
	\end{equation}
	
	In Eq. \ref{eq:update_skewness}, the parameter $\alpha\in (0,2]$ is the scale factor used to control the amplitude of the decrement or increment of average support. Notice that, the term $1/(1+e^{skew(u)})$ has a range of $[0,1]$, which is symmetric at the value of 1/2. So, the term $\alpha\cdot (\frac{1}{1+e^{skew(u)}}-\frac{1}{2})$ has a range of $[-\frac{\alpha}{2},\frac{\alpha}{2}]$ that is symmetric at value of 0. By adjusting $\alpha$ from 0 to 2, the maximum amplitude of the decrement or increment of the original $A(v)$ can be controlled as any value from 0-100\%. For example, if we set $\alpha=1$, then the range of amplitude $[-\frac{1}{2},\frac{1}{2}]$, indicating that a new $A^*(v)$ is up to 50\% higher or lower than the original $A(v)$. We show the effect of $\alpha$ in \S \ref{sec:exp}. Next, we subject Eq. \ref{eq:update_skewness} to Eq. \ref{eq:transition2} to update $\bm{P}$ as follows.
	\begin{equation}
		\label{eq:transition3}
		p_{ij} = \frac{sup(e_{ij})\cdot \frac{A^*(u_j)}{A^*(u_i)}}{\sum_{u_x \in N(u_i)} sup(e_{ix})\cdot \frac{A^*(u_x)}{A^*(u_i)}}
	\end{equation}
	
	\begin{myExample}
		\label{exp:support_distribution}
		Figure \ref{fig:average_support} shows a node having an inflated average support of 22 (but only belongs to a 13-truss). We compute its support skewness by Eq. \ref{eq:skewness} as $skew(u)$ = $18.03>0$, showing it is right-skewed. By setting $\alpha$ = $1$, we decrease its average support as $A^*(u)$ = $22*(1-0.5)$ = $11$, which equals to $k-2$ for $k$ = $13$ (as it belongs to a 13-truss).
	\end{myExample}
	
	\subsection{Optimization with Bound of Trussness}
	\label{sec:rw_op3}
	Compared with an edge's support, its trussness is the most stringent feature to measure a community's cohesiveness. This inspire us to use an edge's trussness to optimize the transition matrix. Since computing an edge's exact trussness in real-time is impractical for the online random walk, we turn to the upper bound of an edge's trussness instead. We next introduce how to compute the upper bound of a node's trussness, and on this basis, how to compute the upper bound of an edge's trussness.
	
	\vspace{0.1cm}
	\begin{myLemma}
		\label{lemma:1}
		Suppose that a node $u$ belongs to a $T_k$, then $u$ has at least $k-1$ adjacent edges with support $sup(e)\geq k-2$.
	\end{myLemma}
	
	\vspace{0.1cm}
	\begin{IEEEproof}
		\label{proof:1}
		Since $u\in T_k$, it has at least one adjacent edge $e_{uv}$ that belongs to $T_k$, where $v\in N(u)$. According to Definition \ref{def:ktruss}, $e_{uv}$ has support $sup(e_{uv})\geq k-2$, which indicates that $u$ and $v$ have at least $k-2$ common neighbors belonging to the same $T_k$. For each common neighbor $w$, the edge $e_{uw}$ still satisfies the constraint of $sup(e_{uw})\geq k-2$. Hence, $u$ has at least $k-1$ adjacent edges with support $sup(e)\geq k-2$.
	\end{IEEEproof}
	
	\vspace{0.1cm}
	Since a node's degree and its adjacent edge's support have a strong correlation with the $k$-truss it belongs to (Lemma \ref{lemma:1}), we define an upper bound of a node's trussness as follows.
	\vspace{0.1cm}
	\begin{myDef}
		\label{def:node_truss_bound}
		\textbf{Upper bound of a node's trussness}. Given a node $u\in V_G$, we define the upper bound of the trussness of $u$, denoted by $\hat{\varphi}(u)$, as the maximum $k$ satisfying Lemma \ref{lemma:1}; that is, $|N(u)|\geq k-1$ and $sup(e_{uv})\geq k-2$ for $\forall v\in N(u)$.
	\end{myDef}
	
	Given an edge $e_{uv}$ between nodes $u,v$, the upper bound of $e_{uv}$'s trussness is determined by $\hat{\varphi}(u)$, $\hat{\varphi}(v)$, and $sup(e_{uv})+2$.
	
	\vspace{0.1cm}
	\begin{myLemma}
		\label{lemma:2}
		Suppose that $sup(e_{uv})+2\geq \min\{\hat{\varphi}(u),\hat{\varphi}(v)\}$, then $e_{uv}$'s trussness is upper bounded by $\min\{\hat{\varphi}(u),\hat{\varphi}(v)\}$; that is, $\phi(e_{uv})\leq \min\{\hat{\varphi}(u),\hat{\varphi}(v)\}$.
	\end{myLemma}
	
	\vspace{0.1cm}
	\begin{IEEEproof}
		\label{proof:2}
		We assume $\min\{\hat{\varphi}(u),\hat{\varphi}(v)\}=k-1$ and $\phi(e_{uv})=k > \min\{\hat{\varphi}(u),\hat{\varphi}(v)\}$. Because $\phi(e_{uv})=k$, $e_{uv}$ belongs to a $k$-truss. So, according to Lemma \ref{lemma:1}, $v,u$ have at least $k-1$ adjacent edges with support $\geq k-2$. Hence, based on Definition \ref{def:node_truss_bound}, we have $\hat{\varphi}(v)$ = $\hat{\varphi}(u)\geq k\Rightarrow \min\{\hat{\varphi}(u),\hat{\varphi}(v)\}\geq k$, which contradicts the assumption $\phi(e_{uv})=k > \min\{\hat{\varphi}(u),\hat{\varphi}(v)\}$. Thus, $\phi(e_{uv})\leq \min\{\hat{\varphi}(u),\hat{\varphi}(v)\}$ holds.
	\end{IEEEproof}
	
	\vspace{0.1cm}
	\begin{myLemma}
		\label{lemma:3}
		Suppose that $sup(e_{uv})+2\leq \min\{\hat{\varphi}(u),\hat{\varphi}(v)\}$, then $e_{uv}$'s trussness is upper bounded by $sup(e_{uv})+2$; that is, $\phi(e_{uv})\leq sup(e_{uv})+2$.
	\end{myLemma}
	
	\vspace{0.1cm}
	\begin{IEEEproof}
		\label{proof:3}
		Suppose $\phi(e_{uv})=k$, then we have $sup(e_{uv})\geq k-2\Rightarrow \phi(e_{uv})\leq sup(e_{uv})+2$. Since $sup(e_{uv})+2\leq \min\{\hat{\varphi}(u),\hat{\varphi}(v)\}$, it is the tightest upper bound of$\phi(e_{uv})$.
	\end{IEEEproof}
	
	\vspace{0.1cm}
	Based on the two aforementioned lemmas, we then define the upper bound of the trussness of an edge as follows.
	
	\vspace{0.1cm}
	\begin{myDef}
		\label{def:edge_truss_bound}
		\textbf{Upper bound of an edge's trussness}. Given an edge $e_{uv}$ with support $sup(e_{uv})$ and two nodes $u,v$ having the upper bound as $\hat{\varphi}(u), \hat{\varphi}(v)$, we define the upper bound of $e_{uv}$'s trussness as $\hat{\phi}(e_{uv})=\min\{sup(e_{uv})+2, \hat{\varphi}(u), \hat{\varphi}(v)\}$.
	\end{myDef}
	
	\vspace{0.1cm}
	We replace $sup(e)$ with $\hat{\phi}(e)$ in Eq. \ref{eq:transition3} to build the connection of transition probability and edge's trussness in Eq. \ref{eq:transition4}.
	\begin{equation}
		\label{eq:transition4}
		p_{ij} = \frac{\hat{\phi}(e_{ij})\cdot \frac{A^*(u_j)}{A^*(u_i)}}{\sum_{u_x \in N(u_i)} \hat{\phi}(e_{ix})\cdot \frac{A^*(u_x)}{A^*(u_i)}}
	\end{equation}
	
	\begin{myExample}
		\label{exp:bound_trussness}
		Recall the example in Figure \ref{fig:average_support}. The upper bound of trussness of node $u_1,u_2,u_8,u_3$ are 6,3,3,6. Thus, we have the upper bound of trussness of edges $\{e_{12},e_{18},e_{13}\}$ as $\hat{\phi}(e_{12})$=$\min\{8+2,3,6\}$=$3$, $\hat{\phi}(e_{18})$=$\min\{1+2,3,6\}$=$3$, and $e_{13}$=$\min\{6,6,6\}$=$6$. We then update the transition probabilities by Eq. \ref{eq:transition4}, e.g., $p_{12}$=$\frac{3\cdot 1.78/2.57}{8 (3\cdot \frac{1}{2.57})+3\cdot \frac{1.78}{2.57}+5 (6\cdot \frac{4}{2.57})}$=$3.58\%$, $p_{13}$=$16.07\%$, and $p_{18}$=$2.01\%$. So, it tends to move from $u_1$ to $\{u_3,\dots, u_7\}$ with the total probability of $80.35\%$ and only has a probability of $19.65\%$ to walk to the left part, which is better compared to the optimization with average support.
	\end{myExample}
	
	\section{Rationality Analysis of the Cohesiveness-aware Transition Matrix}
	\label{sec:bayesian}
Since random walk-based algorithm's effectiveness depends on the cohesiveness-aware transition matrix $\bm{P}$, it's worth discussing the rationality of designing $\bm{P}$ with the selected cohesiveness features, i.e., average support $A(\cdot)$ (Definition \ref{def:average_support} in \S \ref{sec:rw_op1}), skewness of support $skew(\cdot)$ (Eq. \ref{eq:skewness} in \S \ref{sec:rw_op2}), and upper bound of a node's trussness $\hat{\varphi}(\cdot)$ (Definition \ref{def:node_truss_bound} in \S \ref{sec:rw_op3}). Given a query node $q$, we use $T_{k}$ to denote the most cohesive community containing $q$ and $T_{\hat{k}}\subseteq T_k$ ($T_k\setminus T_{\hat{k}}$) is the induced graph of key-members (non-key-members). For $\forall u\in T_k$, it's cohesiveness features are $A(u)$, $skew(u)$, and $\hat{\varphi}(u)$. Intuitively, if these features have a strong correlation with the event that $u$ belongs to $T_{\hat{k}}$ or not, then leverage them to design a cohesiveness-aware transition matrix for random walk is reasonable. More precisely, such a random walk would converge to a stationary distribution of which key-members have greater stationary visiting probabilities than others. 

We first utilize \textit{Bayesian theory} to model the correlation between the cohesiveness features of a node and it's category (i.e., key-member or non-key-members) theoretically, in \S \ref{sec:ba}. Then, we show the correlation results in \S \ref{sec:cr}.


	
	\vspace{-0.1cm}
	\subsection{Correlation Model based on Bayesian Theory}
	\label{sec:ba}
	We aim to use Bayesian theory to compute the probability of a node $u\in T_k$ belongs to key-members $T_{\hat{k}}$ given cohesiveness features $A(u),skew(u),\hat{\varphi}(u)$ as condition (Eq. \ref{eq:bayesian}, \cite{John2013}).	
\begin{equation}
		\label{eq:bayesian}
		\begin{aligned}
			&P\{u \in T_{\hat{k}} \mid X = x_u\} \\
			&= \lim_{\Delta x \to 0} \frac{P\{u \in T_{\hat{k}}\}P\{x_u \leq X \leq x_u + \Delta x \mid u \in T_{\hat{k}}\}}{P\{x_u \leq X \leq x_u + \Delta x\}}
		\end{aligned}
\end{equation}

Here, $X = x_u$ represents the condition given as a node $u$'s features $x_u=\{A(u), skew(u), \hat{\varphi}(u)\}$. The term $P\{u \in T_{\hat{k}}\}$ is the prior knowledge showing the probability of $u$ belongs to $T_{\hat{k}}$, which can be easily computed as the ratio of key-members over all nodes from $T_k$. The term $P\{x_u \leq X \leq x_u + \Delta x\}$ is another prior knowledge showing the probability from all features' \textit{joint distribution} over the entire community $T_k$ within $[x_u, x_u + \Delta x]$. Besides, $P\{x_u \leq X \leq x_u + \Delta x \mid u \in T_{\hat{k}}\}$ is the \textit{class conditional probability} that is computed as the probability from all features' \textit{joint distribution} over $T_{\hat{k}}$ within $[x_u, x_u + \Delta x]$. For the latter two terms, we require to first obtain all features' joint distribution w.r.t. $T_k$ and $T_{\hat{k}}$, respectively, then derive the probability densities $f(x)$ from the joint distributions for computing the cumulative probabilities $\int^{x_u+\Delta x}_{x_u}f(x)dx$, and subject them into Eq. \ref{eq:bayesian}. However, this is non-trivial and we show it from the following observations.

	\begin{figure}
		\setlength{\abovecaptionskip}{0cm}
		\setlength{\belowcaptionskip}{-0.1cm}
		\centering
		\includegraphics[scale=0.46]{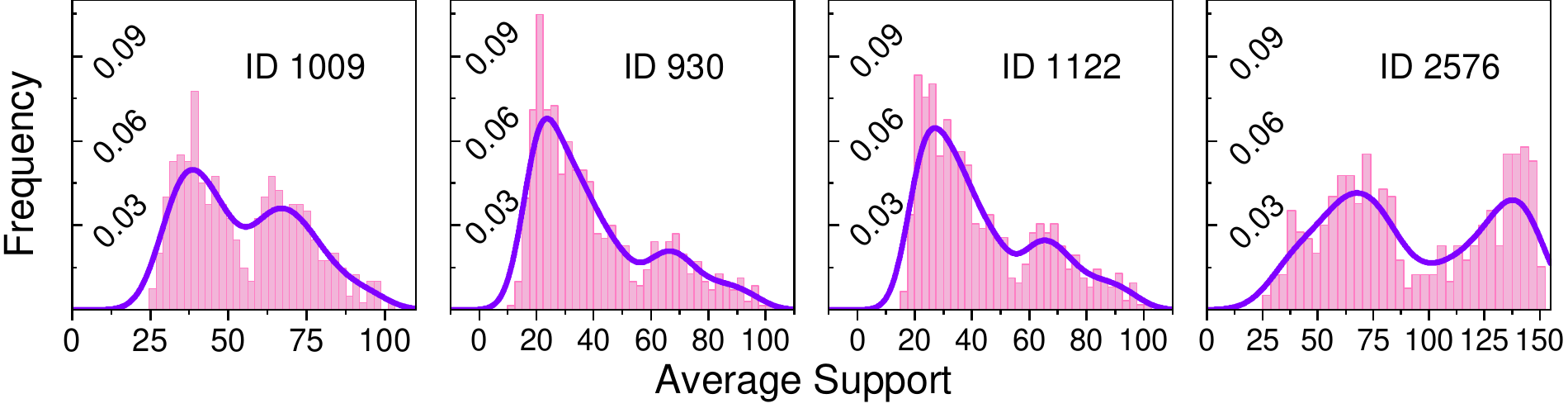}
		\caption{Statistical frequency distribution of $A(\cdot)$ on \texttt{Facebook}}
		\label{fig:bimodal}
	\end{figure}

\vspace{0.1cm}
	\noindent\textbf{Observations.} Figure \ref{fig:bimodal} illustrates the discrete statistical frequency distribution of average support $A(\cdot)$ over $T_k$ on \texttt{Facebook} dataset with three different query nodes (we only provide results for $A(\cdot)$ due to page limit, other features show the similar trend over all datasets). It follows a bimodal distribution including two peaks, each one represents a sub-distribution of $A(\cdot)$ over $T_{\hat{k}}$ and $T_k\setminus T_{\hat{k}}$, respectively. Generally, it's difficult to mathematically model a cohesiveness feature's continuous bimodal distribution from discrete data \cite{Sur2015}, not even the joint distribution of three features. To handle this, we first adopt \textit{Box-Cox Transformation} \cite{Velez2015} to enhance discrete data's normality. Then, we use \textit{Copula Function Fitting} \cite{Escarela2009} to model the joint distributions of three cohesiveness features over $T_{\hat{k}}$ and $T_k\setminus T_{\hat{k}}$, respectively. Finally, we apply \textit{Gaussian Mixture Model} \cite{Reynolds2015} to combine above joint distributions together to obtain the joint distribution over $T_k$.

\vspace{0.1cm}
	\noindent\textbf{Box-Cox Transformation.} For simplicity of discussion, we add superscripts $+$ and $-$ to cohesiveness features to represent the scope where it comes from, key-members $T_{\hat{k}}$ and non-key-members $T_k\setminus T_{\hat{k}}$, respectively, i.e., $A^+$, $A^-$, $skew^+$, $skew^-$, $\hat{\varphi}^+$, and $\hat{\varphi}^-$. Each feature can be viewed as a random variable and we use Box-Cox transformation to enhance its normality as Eq. \ref{eq:boxcox}, where $y$ could be any one of the six random variables, $y_{\rm bct}$ is the transformed value, and $\lambda$ is the transformation factor obtained by parameter estimation \cite{Box1964}. For example, we use $A^+_{\rm bct}(\cdot)$ to indicate the transformed value of the original $A^+(\cdot)$ given the factor $\lambda_{A^+}$ for the feature $A^+$.
	\begin{equation}
		\label{eq:boxcox}
		y_{\rm bct}= 
			\left\{ 
			\begin{aligned}
				&\frac{(y + 1)^{\lambda} - 1}{\lambda} & (\lambda \neq 0)\\
				&log(y + 1) & (\lambda = 0)
			\end{aligned}
			\right .
	\end{equation}
	
Given a transformed random variable, we calculate its mean and variance denoted by $\mu_{y_{\rm bct}}$ and $\sigma^2_{y_{\rm bct}}$. Then, we apply \textit{Copula Function Fitting} to model the joint distribution of cohesiveness features over $T_{\hat{k}}$ (i.e., $y_{\rm bct}\in\{A^+_{\rm bct}, skew^+_{\rm bct}, \hat{\varphi}^+_{\rm bct}\}$) and $T_k\setminus T_{\hat{k}}$ (i.e., $y_{\rm bct}\in\{A^-_{\rm bct}, skew^-_{\rm bct}, \hat{\varphi}^-_{\rm bct}\}$), respectively.
	
\vspace{0.1cm}
	\noindent\textbf{Copula Function Fitting.} \textit{Copula functions} are generally used for multivariate modeling \cite{Escarela2009}, and \textit{Gaussian Copula function} can be used for modeling multivariate Gaussian distribution from multiple unary Gaussian distributions. Let us take the cohesiveness features $y_{\rm bct}\in\{A^+_{\rm bct}, skew^+_{\rm bct}, \hat{\varphi}^+_{\rm bct}\}$ as an example to show the procedure of computing the joint distribution of these features over $T_{\hat{k}}$, via Gaussian Copula function. First, we standardize each transformed random variable as $Z(y_{\rm bct})$ by Eq. \ref{eq:gauss_standradize}. Second, we calculate the covariance between each pair of variables to form the \textit{covariance matrix} of three \textit{marginal distributions} of $Z(A^+_{\rm bct})$, $Z(skew^+_{\rm bct})$, and $Z(\hat{\varphi}^+_{\rm bct})$, denoted by $\Sigma$, and compute the \textit{inverse cumulative distribution function} of each standardized variable, denoted by $\Phi^{-1}(Z(A^+_{\rm bct}))$, $\Phi^{-1}(Z(skew^+_{\rm bct}))$, and $\Phi^{-1}(Z(\hat{\varphi}^+_{\rm bct}))$, respectively.
	\begin{equation}
		\label{eq:gauss_standradize}
		Z(y_{\rm bct}) = \frac{y_{\rm bct} - \mu_{y_{\rm bct}}}{\sigma_{y_{\rm bct}}}
	\end{equation}

Given the inverse cumulative distribution functions $\Phi^{-1}(\cdot)$ of all standardized random variables and the covariance matrix $\Sigma$, we apply Eq. \ref{eq:gaussian_copula} to get the joint distribution of three cohesiveness features over $T_{\hat{k}}$, where $\Phi_{(\mathbf{0}, \Sigma)}$ is the \textit{cumulative distribution function of the multivariate Gaussian distribution} with mean vector $\mathbf{0}=\{0,0,0\}$ and covariance matrix $\Sigma$.
	\begin{equation}
		\label{eq:gaussian_copula}
		\small
		G^+_{\rm joint} = \Phi_{(\mathbf{0}, \Sigma)}(\Phi^{-1}(Z(A^+_{\rm bct})), \Phi^{-1}(Z(skew^+_{\rm bct})), \Phi^{-1}(Z(\hat{\varphi}^+_{\rm bct})))
	\end{equation}
	
	Similarly, the joint distribution $G^-_{\rm joint}$ of three cohesiveness features $y_{\rm bct}\in\{A^-_{\rm bct}, skew^-_{\rm bct}, \hat{\varphi}^-_{\rm bct}\}$ over $T_k \setminus T_{\hat{k}}$, can be obtained following the same aforementioned steps.

\vspace{0.1cm}
	\noindent\textbf{Gaussian Mixture Model (GMM).} We use $f^+(x)$ and $f^-(x)$ to represent the probability density functions of joint distributions $G^+_{\rm joint}$ and $G^-_{\rm joint}$ over $T_{\hat{k}}$ and $T_k\setminus T_{\hat{k}}$, respectively. Then, we apply GMM to compute the probability density function $f(x)$ of the joint distribution over the entire $T_k$, as the weighted sum of $f^+(x)$ and $f^-(x)$ \cite{Reynolds2015} (Eq. \ref{eq:GMM}). The weight assigned on each category is the proportion of nodes belonging to this category, i.e., $P\{ u \in T_{\hat{k}} \}$ and $1-P\{ u \in T_{\hat{k}} \}$. 
	\begin{equation}
		\label{eq:GMM}
		f(x) = P\{ u \in T_{\hat{k}} \} f^+(x) + (1 - P\{ u \in T_{\hat{k}} \} ) f^-(x)
	\end{equation}	

Given the probability density $f(x)$ of the joint distribution over $T_k$ and $f^+(x)$ of the joint distribution over the key-members $T_{\hat{k}}$, we have $P\{x_u \leq X \leq x_u + \Delta x\}=\int^{x_u+\Delta x}_{x_u}f(x)dx$ and $P\{x_u \leq X \leq x_u + \Delta x \mid u \in T_{\hat{k}}\}=\int^{x_u+\Delta x}_{x_u}f^+(x)dx$. By subjecting them into Eq. \ref{eq:bayesian}, we obtain the probability of a node $u$ that belongs to $T_{\hat{k}}$, given the conditions as $u$'s cohesiveness features $A(u),skew(u),\hat{\varphi}(u)$.
	
	\begin{figure}
		\setlength{\abovecaptionskip}{0cm}
		\setlength{\belowcaptionskip}{-0.1cm}
		\centering
		\includegraphics[scale=0.5]{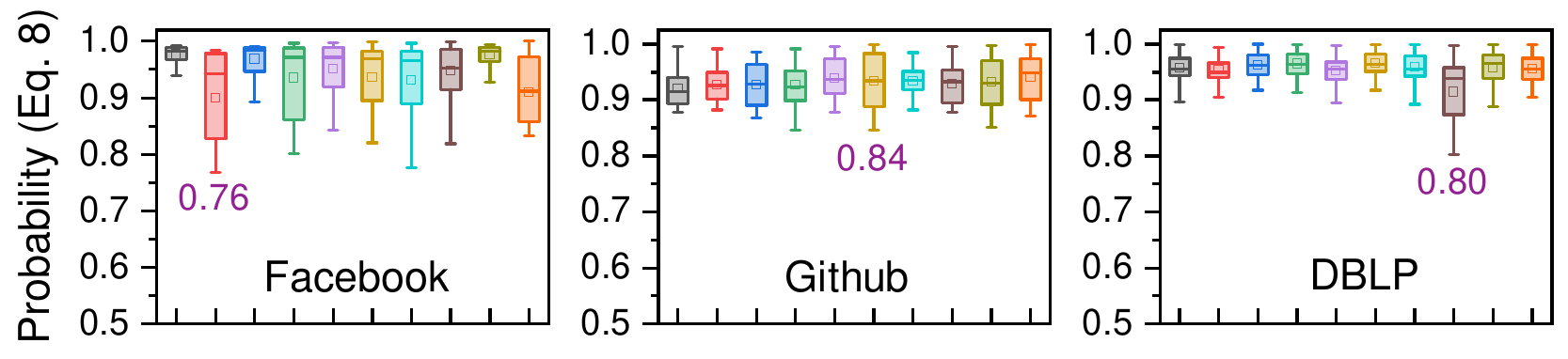}
		\caption{Correlation results of ten queries over three datasets}
		\label{fig:correlation_result}
	\end{figure}
	
	\subsection{Correlation Results}
	\label{sec:cr}
	We apply above correlation model on real-world datasets to estimate key-members' conditional probabilities via Eq. \ref{eq:bayesian}, which are expected to be large values. Due to page limit, we only provide 10 queries' results (corresponds to each point in X-axis) in the form of box plot with min, max, lower-quartile, upper-quartile, and mean probabilities over all key-members (Y-axis) in Figure \ref{fig:correlation_result}, for \texttt{Facebook}, \texttt{GitHub}, and \texttt{DBLP} datasets. For example, the probability on \texttt{Facebook} is at least 76\% (with mean of 90\%), showing that key-members' cohesiveness features and their categories is strongly positively correlated. In a nutshell, using representative cohesiveness features to design transition matrix for our random walk-based algorithm is reasonable theoretically, while our experimental study in \S \ref{sec:exp} shows that it is effective in practice.


	\section{Extension}
	\label{sec:extension}
	We extend our random walk-based algorithms from two aspects. First, we refine the top-$n$ result continuously with a lightweight refinement method (\S \ref{sec:refinement}). Second, we extend it for CKS with multiple query nodes in \S \ref{sec:query_node}.
	
	\setlength{\textfloatsep}{0.1cm}
	\begin{algorithm} [t]
		\small
		\setstretch{0.95}
		\LinesNumbered
		\caption{Top-$n$ result refinement}
		\label{alg:result_rf}
		\KwIn{$S_{{\rm old}}$, $G$} 
		\KwOut{top-$n$ key-members}
		$S_{{\rm new}} \leftarrow $ one-hop neighbors of all nodes in $S_{{\rm old}}$\;
		\For{$\forall u \in S_{{\rm old}}$} 
		{ 
			$|N(u)|=|e_{uv}|$ for $\forall v\in S_{{\rm old}}$\;
		}
		\For{$\forall u \in S_{{\rm new}}$} 
		{ 
			$|N(u)|=|e_{uv}|$ for $\forall v\in S_{{\rm old}}$\;
		}
		$S_{{\rm old}}\cup S_{{\rm new}}\leftarrow$ rank $S_{{\rm old}}\cup S_{{\rm new}}$ by $N(u)$ in descending order\;
		\Return top-$n$ nodes from $S_{{\rm old}}\cup S_{{\rm new}}$\;
	\end{algorithm}
	
	\subsection{Lightweight Result Refinement}
	\label{sec:refinement}
	We use $S_{{\rm old}}$ to denote the top-$n$ key-members returned by a random walk-based algorithm. If $S_{{\rm old}}$ is good enough, then it's expected to contain most of the key-members. Or, we say that the induced graph $G'$ of $S_{{\rm old}}$ has a large overlap with the $T_{\hat{k}}$ to which key-members belong. Since $G'$ is not exactly the same as $T_{\hat{k}}$, it exists at least one ``bad" node in $S_{{\rm old}}$ with neighbors $<\hat{k}-1$ (Lemma \ref{lemma:1}). So, a lightweight method to refine $S_{{\rm old}}$ is to replace these ``bad" nodes in $S_{{\rm old}}$ with other better nodes outside $S_{{\rm old}}$. However, we cannot use $\hat{k}-1$ directly as a lower bound to select these ``bad" nodes, as we do not know the value of $\hat{k}$ in advance. Instead, we use \# neighbors in $S_{{\rm old}}$ of a node $u$ ($N(u)=\{v\mid \forall e_{uv}\in G'\}$) to heuristically measure it's quality. The larger the $|N(u)|$, the better the node $u$.
	
	 Algorithm \ref{alg:result_rf} shows the entire procedure. First, we expand one-hop neighbors of all nodes in $S_{{\rm old}}$ as new candidates for refinement, denoted by $S_{{\rm new}}$ (line 1). Second, for each node $u\in S_{{\rm old}}$, we count its neighbors (lines 2-3). For each node $u\in S_{{\rm new}}$, we count its neighbors (lines 4-5). If a node $u\in S_{{\rm new}}$ has more neighbors in $S_{{\rm old}}$ than that of a node $v\in S_{{\rm old}}$, e.g., $N(u)>N(v)$, then $u$ is more likely to be a better node than $v$. So, we rank all nodes in $S_{{\rm old}}\cup S_{{\rm new}}$ by $N(\cdot)$ in descending order, and return the top-$n$ nodes as the refined key-members. 
	
	\begin{figure}
		\setlength{\abovecaptionskip}{0.1cm}
		\setlength{\belowcaptionskip}{-0.1cm}
		\centering
		\hspace{-0.4cm}
		\includegraphics[scale=0.32]{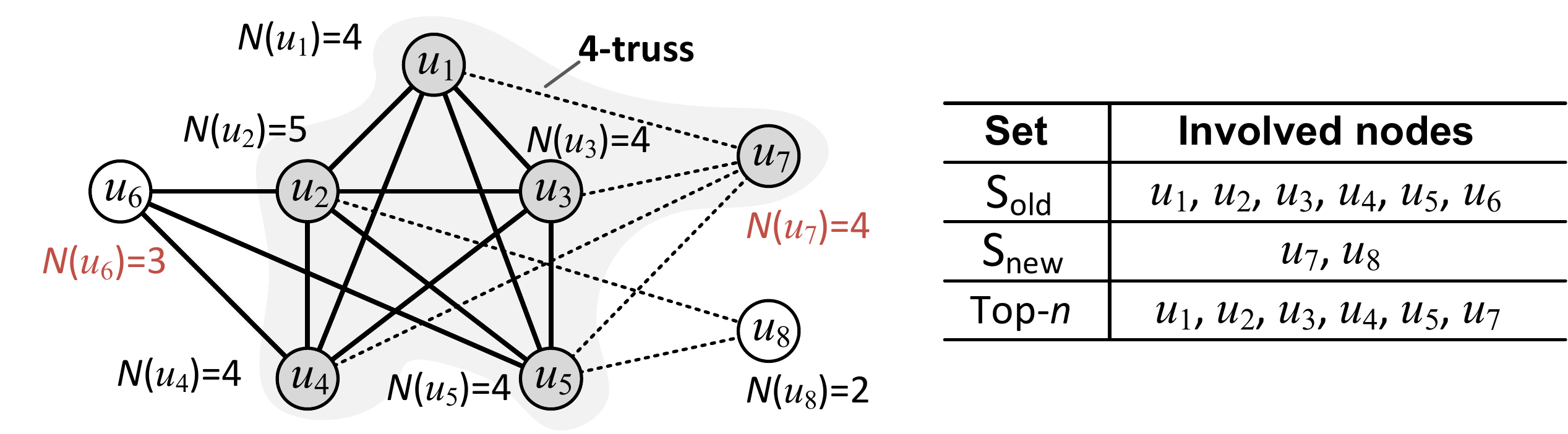}
		\caption{Top-$n$ key-members refinement}
		\label{fig:refinement}
	\end{figure}
	
	\vspace{0.1cm}
	\begin{myExample}
		\label{exp:refinement}
		Figure \ref{fig:refinement} illustrates an example of lightweight refinement. Given the top-6 nodes $S_{{\rm old}}=\{u_1,\dots,u_6\}$, which forms a 3-truss. We expand two new candidates $u_7,u_8$, then count the number of neighbors (in $S_{{\rm old}}$) of all nodes and replace $u_6$ with $u_7$, because $N(u_7)>N(u_6)$. Finally, we return new top-$6$ results that can form a more cohesive 4-truss.
	\end{myExample}
	
	\vspace{0.1cm}
	In \S \ref{sec:exp}, we show that the precision can reach to 97\% on average after two iterations of Algorithm \ref{alg:result_rf}. For example, precision reaches to nearly 100\% for \texttt{GitHub} and \texttt{DBLP}. Besides, the overhead is modest, e.g., extra 2 ms on average for \texttt{GitHub} to improve the precision from 89.3\% to 99.3\%.
	\subsection{Support to Multiple Query Nodes}
	\label{sec:query_node}
	Using multiple query nodes to find key-members is common in some real applications, such as searching for key-members of criminal gangs by providing a group of suspects. According to the original definition of CKS, key-members for the case of multiple query nodes would be those nodes in the same community with all query nodes and have maximum trussness. Given a group of query nodes $Q=\{q_1,\dots, q_n\}$, we extend our random walk-based algorithms as follows. First, we extract the $m$-bounded subgraph for each query node $q_i\in Q$ from the original graph $G$, denoted by $G_{q_i}$. Second, we take the union of these subgraphs as the $m$-bounded subgraph of $Q$, denoted by $G_Q$. After that, we can apply an arbitrary algorithm from \S \ref{sec:rw_basic}-\ref{sec:rw_op3} to find the key-members in $G_Q$. 
\begin{equation}
\label{eq:union_subgraph}
G_Q=G_{q_1}\cup G_{q_2}\dots \cup G_{q_n}
\end{equation}

Since we do not terminate the random walk when it converges (see \S \ref{sec:rw_basic}), walking from different query node would generate different stationary distribution. Technically, we need to perform $|Q|$ times random walk from each node in $Q$, then return the final top-$n$ nodes by considering all $|Q|$ stationary distributions. Fortunately, the difference in $|Q|$ stationary distributions is too small to affect the final result if we set large enough iterations $r$, e.g., 150, for matrix multiplication. So, in our implementation, we only randomly select one of $|Q|$ query nodes to perform the random walk.

\vspace{0.1cm}
\noindent\textbf{Remarks.} The structure between $|Q|$ query nodes is sometimes important for CKS. If we can estimate the lower bound of trussness of edges among $|Q|$ query nodes, denoted by $\check{\phi}(Q)$, then we can assign a smaller transition probability for those edges with the upper bound of trussness $\hat{\phi}(e)< \check{\phi}(Q)$. We keep this as an interesting open problem for future work.
	
	\begin{figure*}
\setlength{\abovecaptionskip}{0.05cm}
 \hspace{-0.6cm}
 \begin{minipage}{0.26\linewidth}
    \includegraphics[scale=0.4]{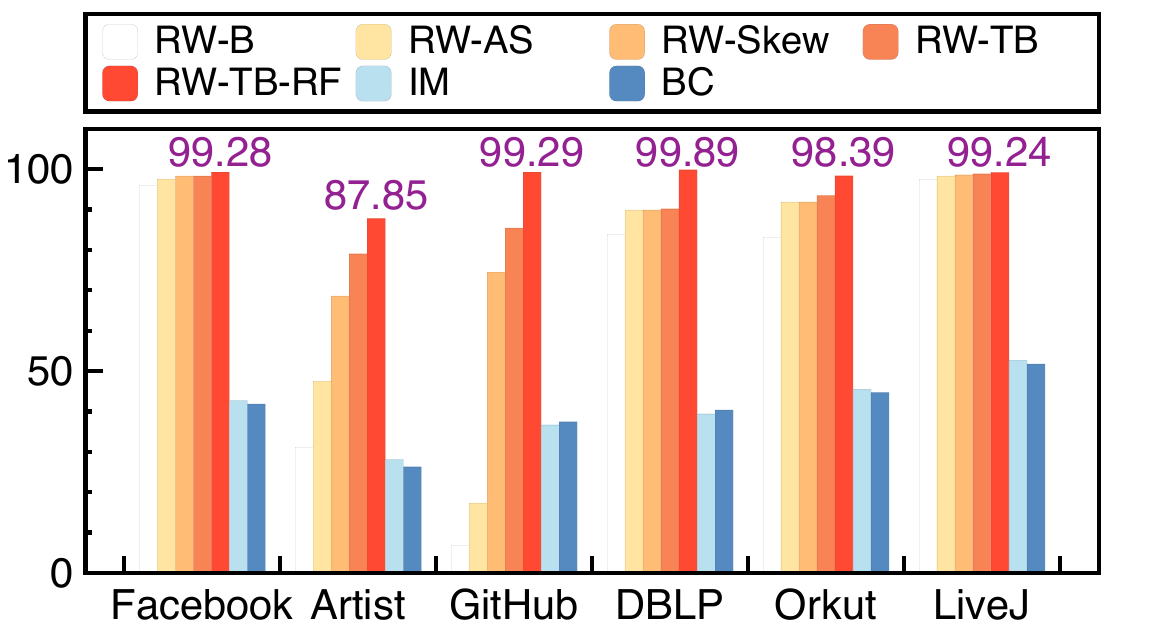}
    \vspace{-0.4cm}
    \caption{Precision (\%)}
    \label{fig:effect_efficiency1}
  \end{minipage}
  \hspace{-0.2cm}
  \begin{minipage}{0.26\linewidth}
    \includegraphics[scale=0.4]{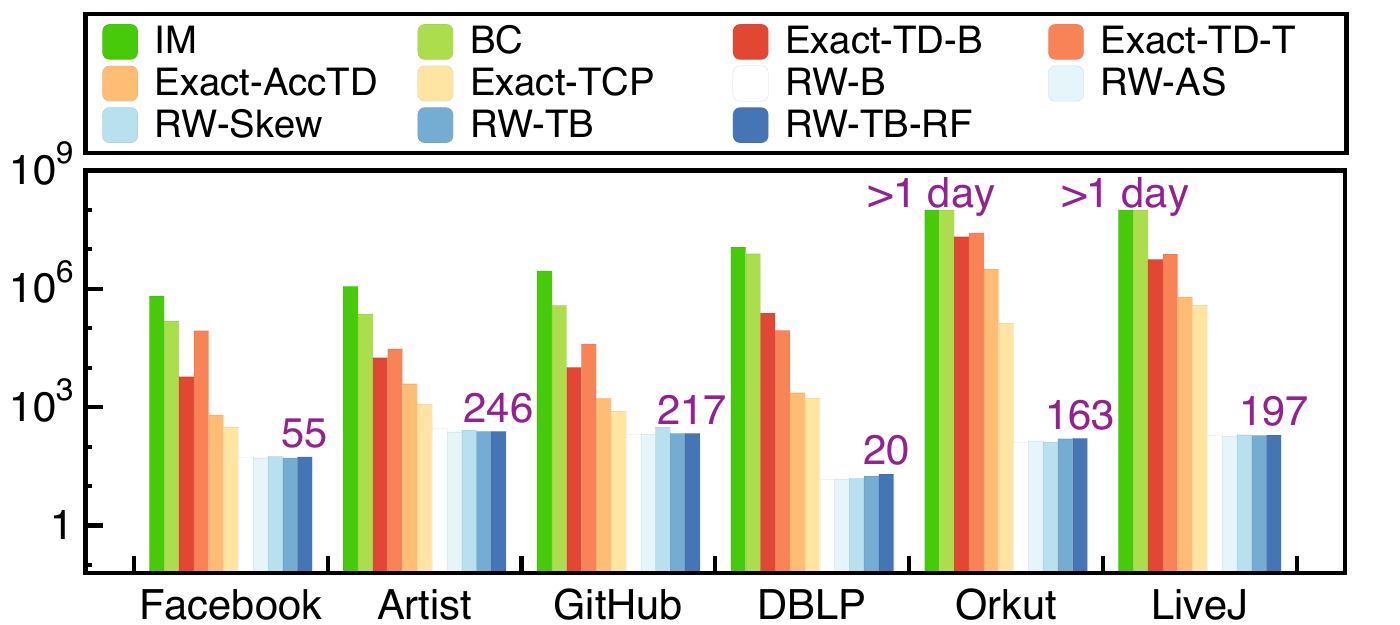}
    \vspace{-0.4cm}
    \caption{Response time (ms)}
    \label{fig:effect_efficiency2}
  \end{minipage}
  \begin{minipage}[b]{0.58\textwidth}
    \centering
    \tabcaption{Effectiveness: diameter (DM) and density (DS)}
    		\scalebox{0.52}{
			\begin{tabular}{c||cc||cc||cc||cc||cc||cc}
				\multirow{2}{*}{\textbf{Methods} $\downarrow$}      & \multicolumn{2}{c||}{Facebook}           & \multicolumn{2}{c||}{GitHub}             & \multicolumn{2}{c||}{Artist}             & \multicolumn{2}{c}{DBLP}              & \multicolumn{2}{c||}{Orkut}             & \multicolumn{2}{c}{LiveJournal}   \\
				& \multicolumn{1}{c|}{DM} & DS & \multicolumn{1}{c|}{DM} & DS & \multicolumn{1}{c|}{DM} & DS & \multicolumn{1}{c|}{DM} & DS & \multicolumn{1}{c|}{DM} & DS & \multicolumn{1}{c|}{DM} & DS \\ \hline\hline
				Exact-TD& \multicolumn{1}{c|}{2.0}      & 64.7   & \multicolumn{1}{c|}{2.0}      & 17.7   & \multicolumn{1}{c|}{2.0}      & 13.3   & \multicolumn{1}{c|}{2.0}      & 56.5  & \multicolumn{1}{c|}{2.0}      &  55.35  & \multicolumn{1}{c|}{2.0}      &   182.64  \\ \hline\hline
				RW-B    & \multicolumn{1}{c|}{2.04}     & 63.4   & \multicolumn{1}{c|}{3.64}     & 8.9    & \multicolumn{1}{c|}{2.16}      & 7.8    & \multicolumn{1}{c|}{2.5}      & 48.5 & \multicolumn{1}{c|}{2.03}      &  54.23  & \multicolumn{1}{c|}{2.02}      & 181.43  \\ \hline
				RW-AS   & \multicolumn{1}{c|}{2.02}     & 63.7   & \multicolumn{1}{c|}{3.22}     & 9.8    & \multicolumn{1}{c|}{2.0}     & 8.6    & \multicolumn{1}{c|}{2.1}      & 52.4  & \multicolumn{1}{c|}{2.02}      &  54.34  & \multicolumn{1}{c|}{2.0}      &  181.79  \\ \hline
				RW-Skew & \multicolumn{1}{c|}{2.0}      & 64.3   & \multicolumn{1}{c|}{3.18}     & 10.1   & \multicolumn{1}{c|}{2.03}     & 10.4   & \multicolumn{1}{c|}{2.08}     & 53.1  & \multicolumn{1}{c|}{2.0}      &  55.02  & \multicolumn{1}{c|}{2.0}      &  182.42   \\ \hline
				RW-TB   & \multicolumn{1}{c|}{2.0}      & 64.3   & \multicolumn{1}{c|}{2.94}     & 10.9   & \multicolumn{1}{c|}{2.0}      & 10.7   & \multicolumn{1}{c|}{2.08}     & 54.1  & \multicolumn{1}{c|}{2.0}      &  55.25  & \multicolumn{1}{c|}{2.0}      &  182.59  \\ \hline
				RW-TB-RF& \multicolumn{1}{c|}{2.0}      & 64.7   & \multicolumn{1}{c|}{2.08}     & 16.0   & \multicolumn{1}{c|}{2.0}      & 13.3   & \multicolumn{1}{c|}{2.0}      & 56.3  & \multicolumn{1}{c|}{2.0}      &  55.34  & \multicolumn{1}{c|}{2.0}      & 182.64 \\ \hline\hline
				IM   & \multicolumn{1}{c|}{3.0}      & 0.71   & \multicolumn{1}{c|}{3.0}     & 0.32   & \multicolumn{1}{c|}{4}      & 0.23   & \multicolumn{1}{c|}{3.0}     & 2.67  & \multicolumn{1}{c|}{3.0}      &  0.98  & \multicolumn{1}{c|}{3.0}      &  4.36  \\ \hline
				BC & \multicolumn{1}{c|}{4.0}      & 0.05   & \multicolumn{1}{c|}{4.0}     & 0.03   & \multicolumn{1}{c|}{4.0}      & 0.02   & \multicolumn{1}{c|}{4.0}      & 0.64  & \multicolumn{1}{c|}{4.0}      &  0.81  & \multicolumn{1}{c|}{4.0}      & 1.33
			\end{tabular}
			}
    \label{tab:density}
  \end{minipage}
  \vspace{-0.6cm}
\end{figure*}
	
	\section{Experiments}
	\label{sec:exp}
	We evaluate (1) effectiveness (\S \ref{sec:effect}), (2) efficiency (\S \ref{sec:efficiency}), (3) case study (\S \ref{sec:case}), (4) parameter sensitivity (\S \ref{sec:para}), and (5) overhead of the refinement method (\S \ref{sec:refine}). Our code and datasets were provided in \cite{code}. All experiments were run on a 3.7 GHZ, 128 GB memory Linux server.
	
	\subsection{Experimental Setup}
	
	\noindent \textbf{Datasets.} We used six real-world datasets with statistics shown in Table \ref{tab:datasets} (e.g., maximum degree $d_{\rm max}$, support $s_{\rm max}$, and trussness $k_{\rm max}$). Aritist \cite{Rozemberczki2019} maintains nodes as the blue verified Facebook pages with artist category, and edges are mutual likes among them. Facebook \cite{McAuley2012} is an anonymous dataset containing friend lists of users. GitHub \cite{Rozemberczki2021} involves the developers in GitHub who have starred at least 10 repositories and edges are mutual follower relationships between them. DBLP \cite{Yang2015} provides relationships among authors, papers, venues, etc. We constructed a homogeneous co-authorship network where two authors are connected if they have co-authored at least one paper. Orkut \cite{Orkut2022} is a social network extracted from \url{Orkut.com}. LiveJournal \cite{LiveJournal2022} is a free online blogging community where users declare friendship each other.
	
	\vspace{0.1cm}
	\noindent \textbf{Queries.} For each dataset, we randomly selected 5000 query nodes to perform CKS and report the average effectiveness and efficiency results. We generated the ground truth key-members for each query by running exact algorithm presented in \S \ref{sec:exact}.
	
	\setlength{\textfloatsep}{0cm}
	\begin{table}
		\setlength{\abovecaptionskip}{0cm}
		\centering
		\caption{Statistics of datasets}
		\label{tab:datasets}
		\scalebox{0.68}{
			\begin{tabular}{c||c|c|c|c|c|c}
				\textbf{Datasets} $\downarrow$ & \# \textbf{Nodes}           & \# \textbf{Edges}	& \textbf{$d_{\rm max}$} & \textbf{$s_{\rm max}$} & \textbf{$k_{\rm max}$}	&	\# Triangles \\ \hline \hline
				Artist		&	50,515	&	819,306	&	1,469	&	735	&	23	&	2,273,700	\\ \hline
				Facebook (FB)		&	4,039	&	88,234		&	1,045	&	293	&	97	&	1,612,010	\\ \hline
				GitHub		&	37,700	&	289,003	&	9,458	&	2,411	&	24	&	523,809	\\ \hline
				DBLP 		&	317,080	&	1,049,866	&	343	&	213	&	114	&	2,224,385	 \\ \hline
				Orkut		& 3,072,441 & 117,185,082 & 33,313 & 9,145 & 78 & 627,584,176 \\ \hline
				LiveJournal (LiveJ) & 3,997,962 & 34,681,189 & 14,815 & 1,393 & 352 & 177,820,130 \\
			\end{tabular} 
		}
	\end{table}
	
	\vspace{0.1cm}
	\noindent \textbf{Metrics.} We used the precision, recall, $F1$-score to measure the accuracy of returned key-members w.r.t. the ground truth. Besides, we used diameter \cite{Huang2015} and density \cite{Wu2015} of a graph to evaluate the closeness of the induced graph of key-members, which is the complement to structure cohesiveness of $k$-truss. We used the response time for efficiency evaluation.
	
	
	\vspace{0.1cm}
	\noindent \textbf{Comparing methods.} We compared with four exact algorithms: (1) Exact-TD-B is established atop TD-bottomup \cite{Wang2012}, (2) Exact-TD-T based on TD-topdown \cite{Wang2012}, (3) Exact-AccTD is extended from AccTD \cite{Che2020}, and (4) Exact-TCP-Index (shorten as Exact-TCP) based on TCP-Index \cite{Huang2014}. We implemented four random walk-based algorithms (\S \ref{sec:rw_basic}-\S \ref{sec:rw_op3}): the basic (5) RW-B, (6) RW-AS with optimization of average support, (7) RW-Skew with optimization of support skewness, and (8) RW-TB with optimization of trussness bound. We integrated the refinement method (\S \ref{sec:refinement}) with RW-TB to form (9) RW-TB-RF. Besides, we compared with two critical nodes identification methods: (10) Influence maximization (IM) \cite{Wang2017} and (11) Betweenness centrality (BC) \cite{Mumtaz2017}.
	
	\vspace{0.1cm}
	\noindent \textbf{Parameters.} The default parameters are: $m$-bounded subgraph of $m$ = $2$, iterations $r$ = $150$, scale factor $\alpha$ = $1$ for skewness, top-$n$ key-members of $n$ = $|$ground truth$|$, and $|Q|$ = $1$.
	
	\subsection{Effectiveness Evaluation}
	\label{sec:effect}
	\noindent\textbf{Precision.} Figure \ref{fig:effect_efficiency1} shows the precision results. Since we set $n$ = $|$ground truth$|$, the precision, recall, and $F1$-scores are equalized. So, we only provide the precision. In \S \ref{sec:para}, we will show the effect of $n$ on three metrics. Since exact algorithms have 100\% precision, we omit them from Figure \ref{fig:effect_efficiency1}. RW-B performs the worst among random walk-based algorithms (66\% on average), but for Facebook and LiveJournal, it has at least 96\% precision. This indicates that it is feasible to use edge support to design transition matrix, but it is not enough to get a good result. The average precision is improved to 74\%, 86\%, and 90\% by using RW-AS, RW-Skew, and RW-TB. The improvement is obvious on GitHub and Artist, as they contains more nodes having skewed support distribution than others. So, it's useful to fine-tune the average support by considering the support skewness. Besides, RW-TB-RF with additional two iterations of refinement achieves 97\% precision on average (we show precision on the top of bars), and some results reach to nearly 100\%, showing that our refinement method is effective. 


\begin{figure*}
  \setlength{\abovecaptionskip}{-0.1cm}
  \hspace{-0.4cm}
  \subfigure[Effect of $m$ on precision]{%
  \begin{minipage}{0.18\linewidth}
    \setlength{\abovecaptionskip}{0.1cm}
    \includegraphics[scale=0.31]{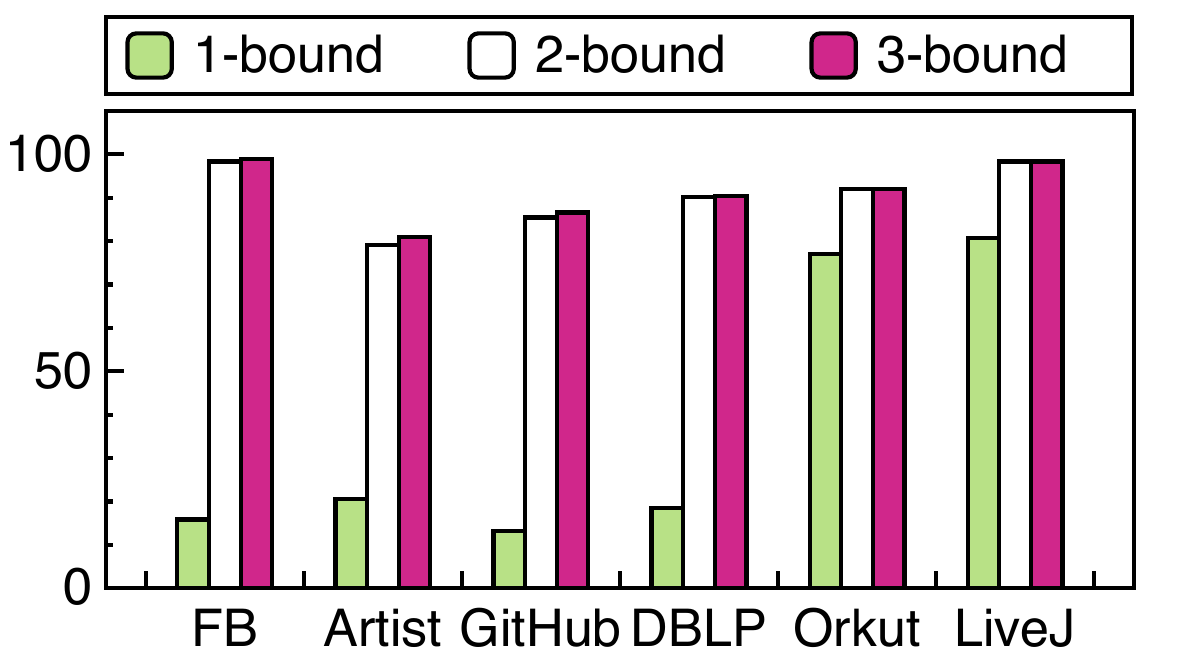}
    \vspace{-0.3cm}
  \end{minipage}
  }
  \hspace{0.1cm}
  \subfigure[Effect of $m$ on runtime]{%
  \begin{minipage}{0.18\linewidth}
    \setlength{\abovecaptionskip}{0.1cm}
    \includegraphics[scale=0.31]{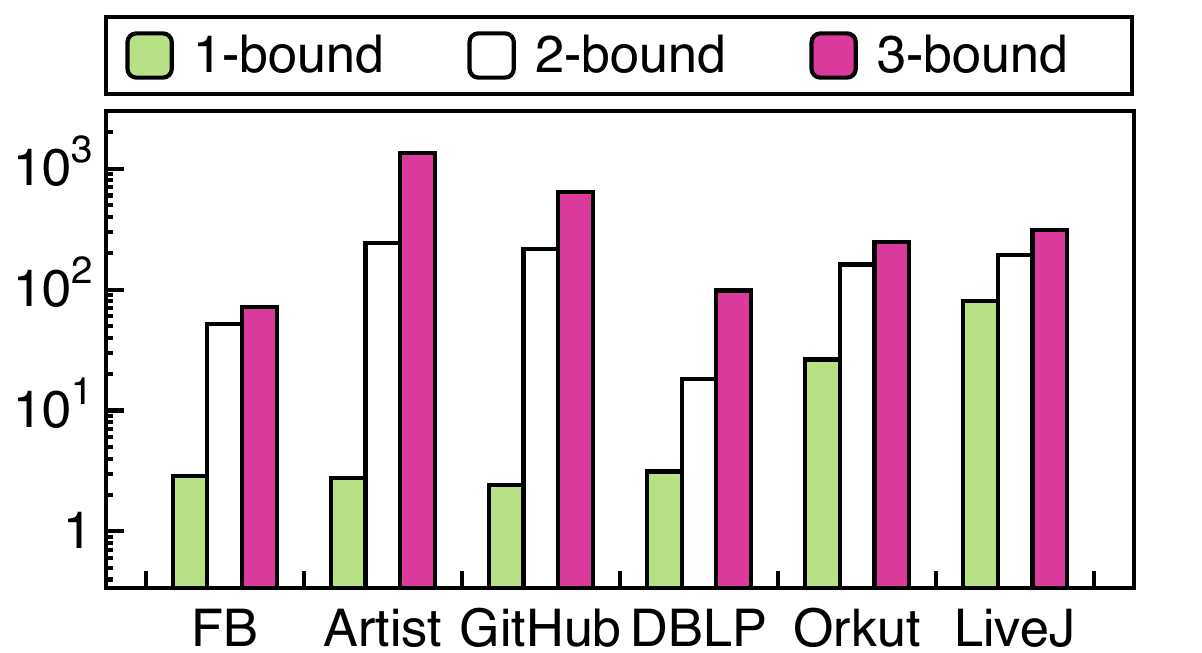}
    \vspace{-0.3cm}
  \end{minipage}
  }
 \hspace{0.1cm}
  \subfigure[Effect of $r$ on precision]{%
  \begin{minipage}{0.18\linewidth}
    \setlength{\abovecaptionskip}{0.1cm}
    \includegraphics[scale=0.31]{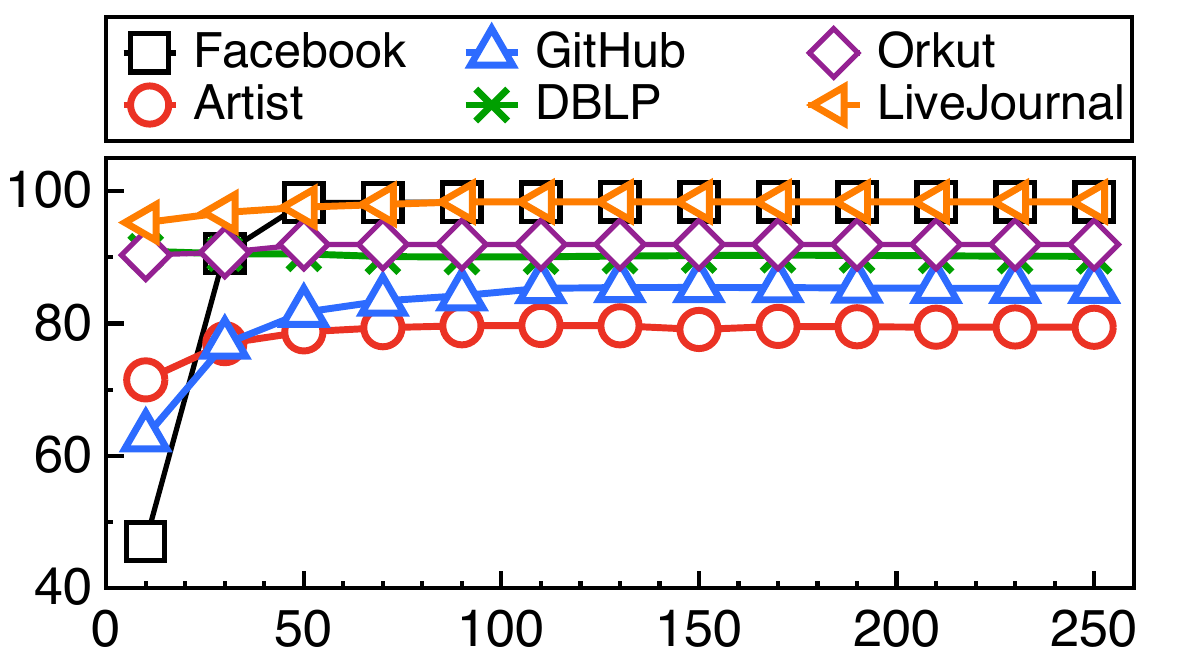}
    \vspace{-0.3cm}
  \end{minipage}
  }
  \hspace{0.1cm}
  \subfigure[Effect of $r$ on runtime]{%
  \begin{minipage}{0.18\linewidth}
    \setlength{\abovecaptionskip}{0.1cm}
    \includegraphics[scale=0.31]{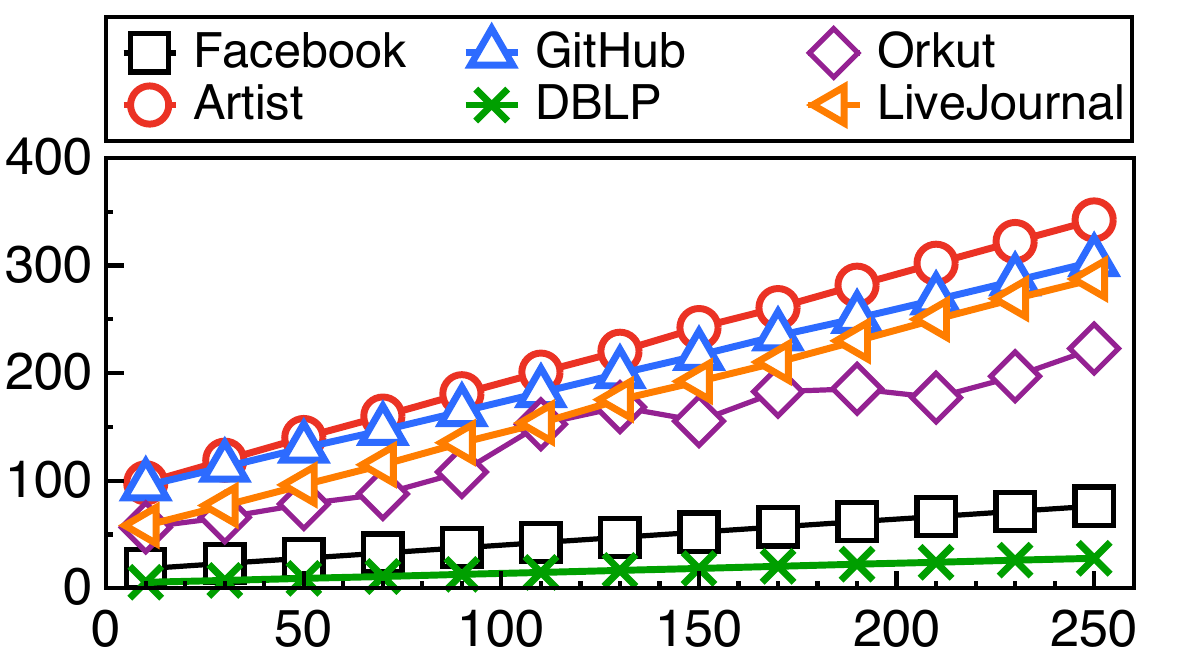}
    \vspace{-0.3cm}
  \end{minipage}
  }
  \hspace{0.1cm}
  \subfigure[Effect of $\alpha$ on precision]{%
  \begin{minipage}{0.18\linewidth}
    \setlength{\abovecaptionskip}{0.1cm}
    \includegraphics[scale=0.31]{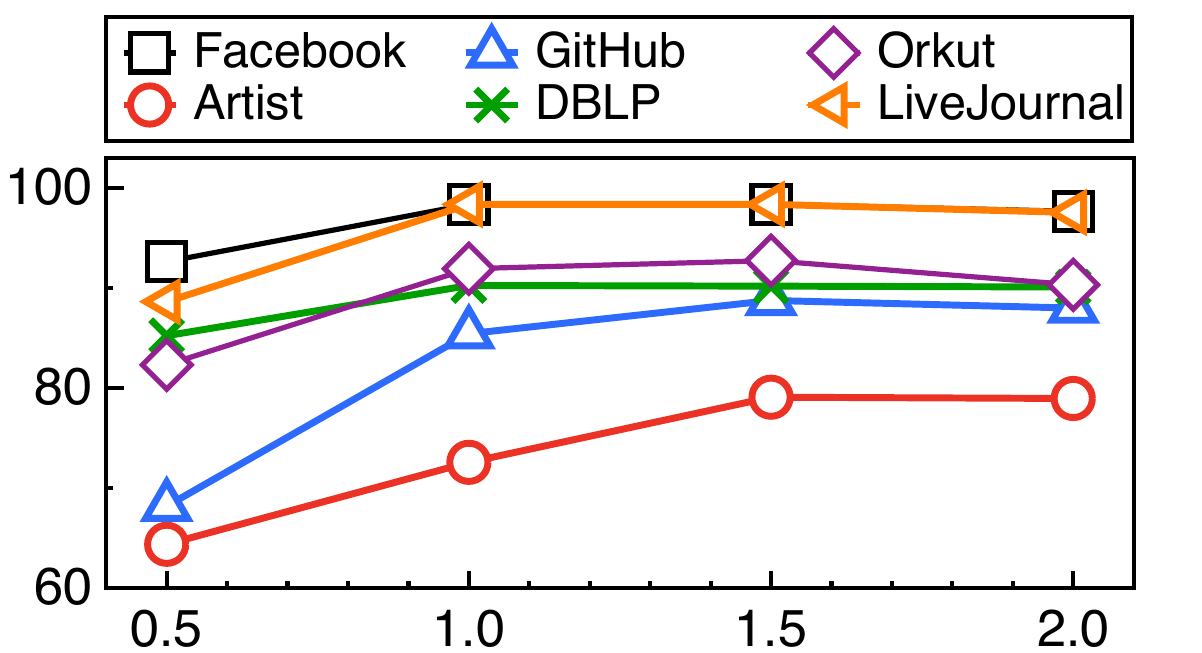}
    \vspace{-0.3cm}
  \end{minipage}
  }
  \caption{Effect of the $m$-bounded subgraph, random walk iterations $r$, and scale factor $\alpha$ on precision (\%) and runtime (ms)}
  \label{fig:parameter}
  \vspace{-0.6cm}
\end{figure*}		
	
	\vspace{0.1cm}
	\noindent\textbf{Diameter and density.} Table \ref{tab:density} shows the diameter and density of the induced graph formed by key-members. Note that, key-members returned by exact algorithms form the most cohesive $k$-truss with the smallest diameter and largest density. For ours, the diameter decreases (density increases) as we use a better RW algorithm, and RW-TB-RF perform the best, which is very close to the ground truth. This proves that our key-members are closely connected and have a large overlap with the ground truth, thus resulting in a good precision. IM and BC do not consider the cohesiveness of critical nodes, thus leading to a result with larger diameter and a smaller density.

	\vspace{-0.1cm}
	\subsection{Efficiency Evaluation}
	\label{sec:efficiency}
	In Figure \ref{fig:effect_efficiency2}, exact algorithms are time-consuming as they rely on the heavyweight truss-decomposition. Exact-AccTD and Exact-TCP-Index are more efficient than other exact algorithms as they are benefit from well-designed data structures or index. However, it introduce additional space overhead for maintaining index, e.g., 3 GB index is required for 1 GB Orkut dataset.  Our random walk-based algorithms outperform exact algorithms (we show RW-TB-RF's runtime on the top of bars), e.g., ours are at least 3.8X and 521X on average faster than others over all datasets. This is because ours do not maintain accurate truss information in runtime, but only leverage the random walk to visit nodes that are most likely to belong to the most cohesive $k$-truss. The overhead introduced by refinement method is modest, e.g., extra 2 ms on average for GitHub to improve the precision from 89.3\% to 99.3\% (see \S \ref{sec:refine}).
	
\begin{figure}[h]
		\setlength{\abovecaptionskip}{0.1cm}
		\setlength{\belowcaptionskip}{0cm}
		\centering
		\includegraphics[scale=0.32]{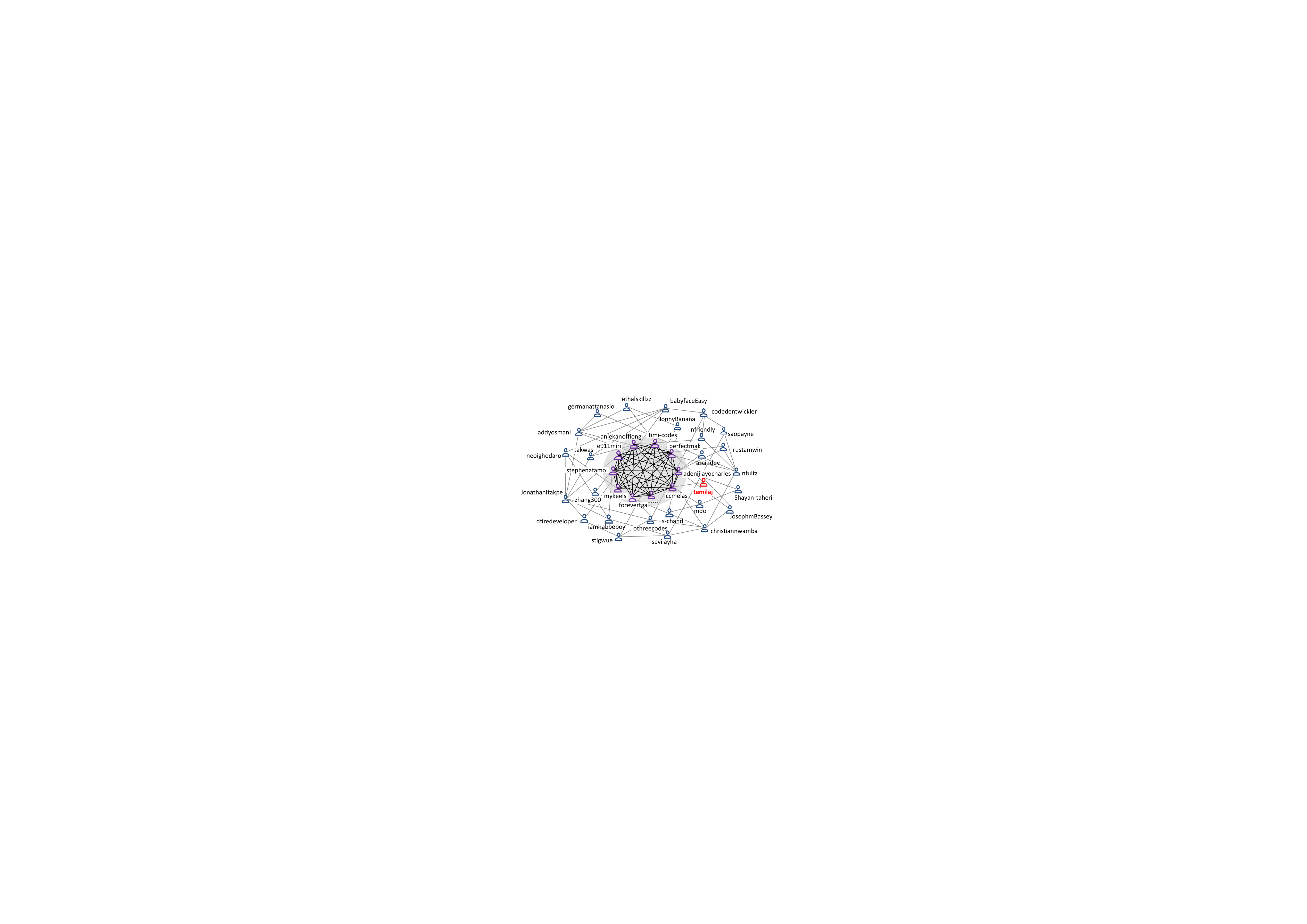}
		\caption{A case study on GitHub (RW-TB)}
		\label{fig:case_study}
	\end{figure}		
	
	\vspace{-0.1cm}
	\subsection{Case Study}
	\label{sec:case}
	We run a case study on GitHub by RW-TB ($q$ = temilaj with ID 20053). Figure \ref{fig:case_study} illustrates the returned key-members and their relationships (gray area). We only provide 10 out of 28 key-members and other 25 users due to the page limit. These key-members participate in the densest subgraph (a 24-truss), showing that they are closely connected. Each pairwise key-members have at least 22 common followers, most of them are full stack web developer and share with the same skills, e.g., React, AI, Go, JS. Moreover, these key-members belong to the most cohesive community that contains the query node (a 20-truss). By performing CKS, the user with ID 20053 can find the key circle she interested in (i.e., the gray area) and expand her social circle in GitHub by following them. Our solution can find these key-members because the random walk is guided toward the most cohesive 24-truss with a higher probability. 

	\vspace{-0.1cm}
	\subsection{Parameter Sensitivity}
	\label{sec:para}
\begin{figure}
  \vspace{-0.1cm}
  \setlength{\abovecaptionskip}{-0.1cm}
  \setlength{\belowcaptionskip}{0cm}
   \hspace{-0.3cm}
  \subfigure[Facebook]{
  \begin{minipage}{0.26\linewidth}
    \includegraphics[scale=0.31]{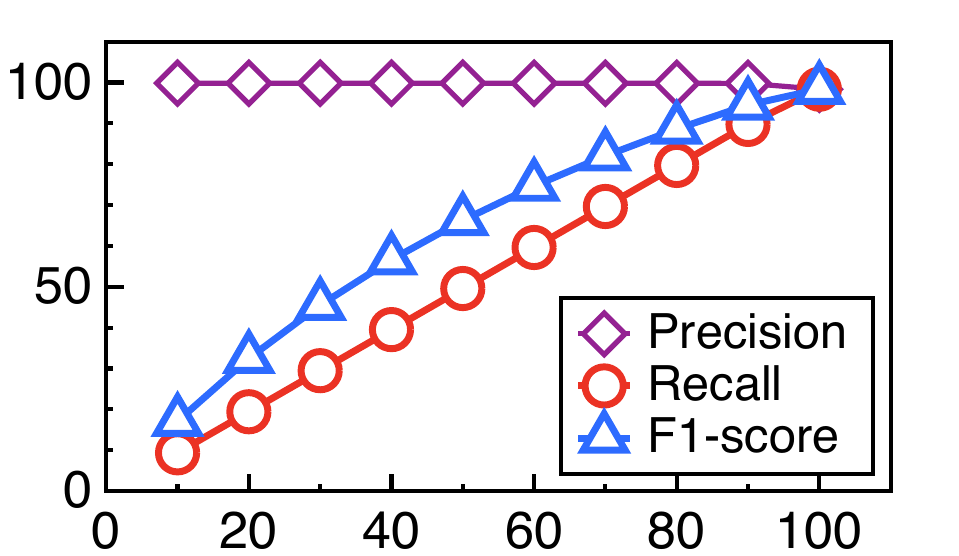}
    \vspace{-0.3cm}
  \end{minipage}
  }
  \hspace{0.1cm}
  \subfigure[Artist]{
  \begin{minipage}{0.26\linewidth}
    \includegraphics[scale=0.31]{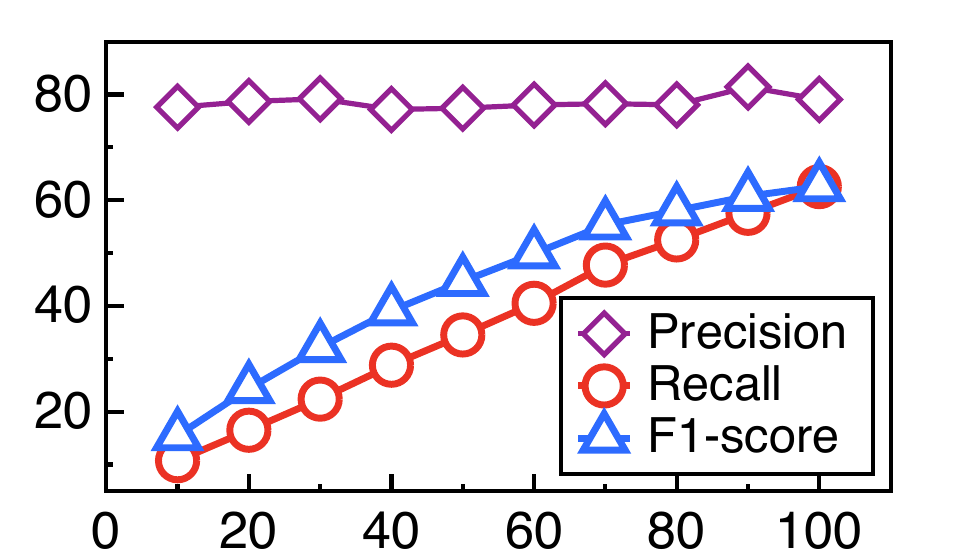}
    \vspace{-0.3cm}
  \end{minipage}
  }
  \hspace{0.1cm}
  \subfigure[GitHub]{
  \begin{minipage}{0.26\linewidth}
    \includegraphics[scale=0.31]{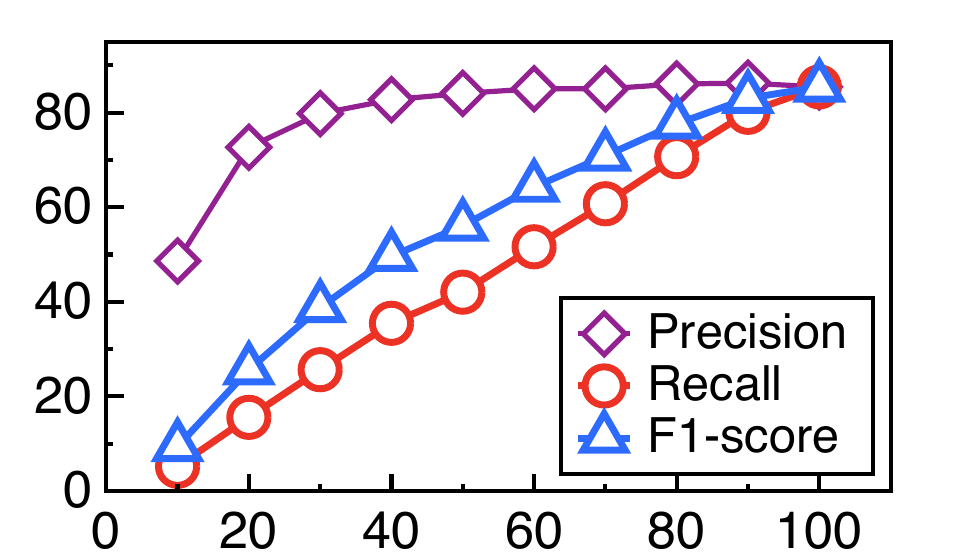}
    \vspace{-0.3cm}
  \end{minipage}
  }
  
  \vspace{-0.2cm}
  \hspace{-0.3cm}
  \subfigure[DBLP]{
  \begin{minipage}{0.26\linewidth}
    \includegraphics[scale=0.31]{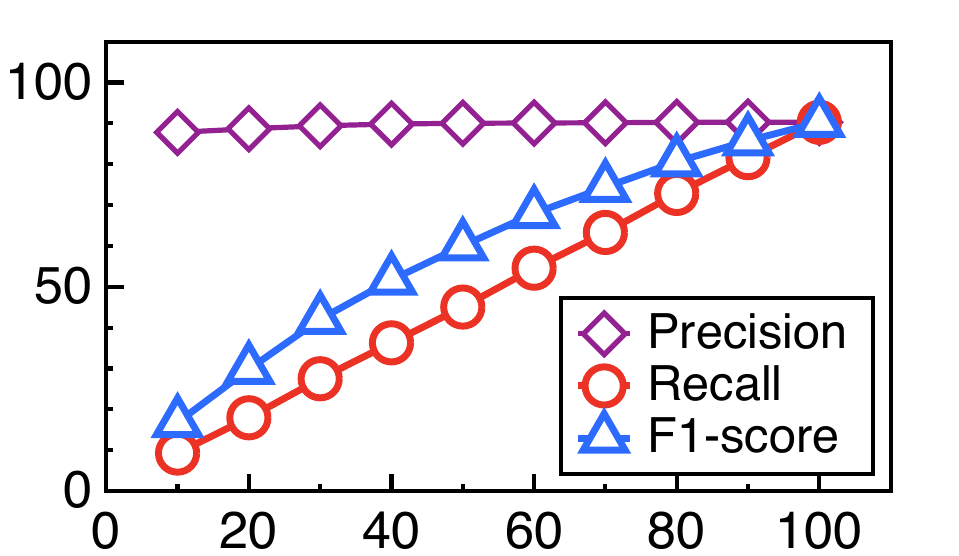}
    \vspace{-0.3cm}
  \end{minipage}
  }
  \hspace{0.1cm}
  \subfigure[Orkut]{
  \begin{minipage}{0.26\linewidth}
    \includegraphics[scale=0.31]{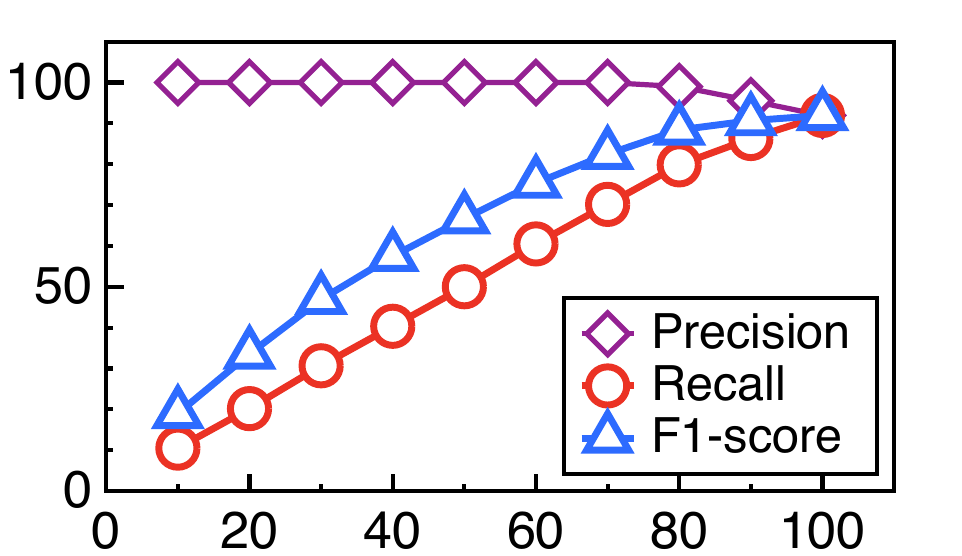}
    \vspace{-0.3cm}
  \end{minipage}
  }
  \hspace{0.1cm}
  \subfigure[LiveJournal]{
  \begin{minipage}{0.26\linewidth}
    \includegraphics[scale=0.31]{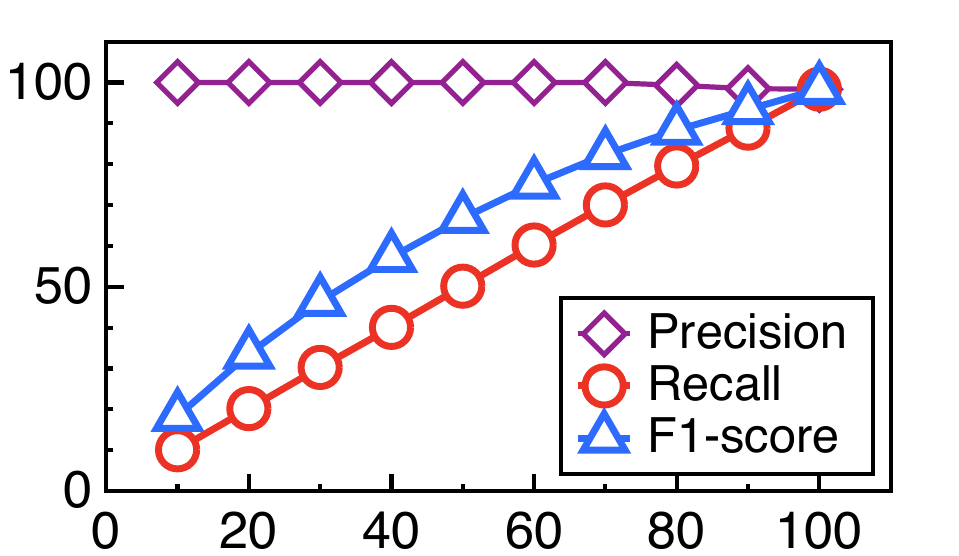}
    \vspace{-0.3cm}
  \end{minipage}
  }
  \caption{Effect of the top-$n$ nodes ($n=f\cdot|$ground truth$|$)}
  \label{fig:topn}
  \vspace{-0.1cm}
\end{figure}

\begin{figure}
  \setlength{\abovecaptionskip}{-0.1cm}
  \setlength{\belowcaptionskip}{0cm}
  \hspace{-0.4cm}
  \subfigure[Effect on precision (\%)]{
  \begin{minipage}{0.45\linewidth}
    \includegraphics[scale=0.36]{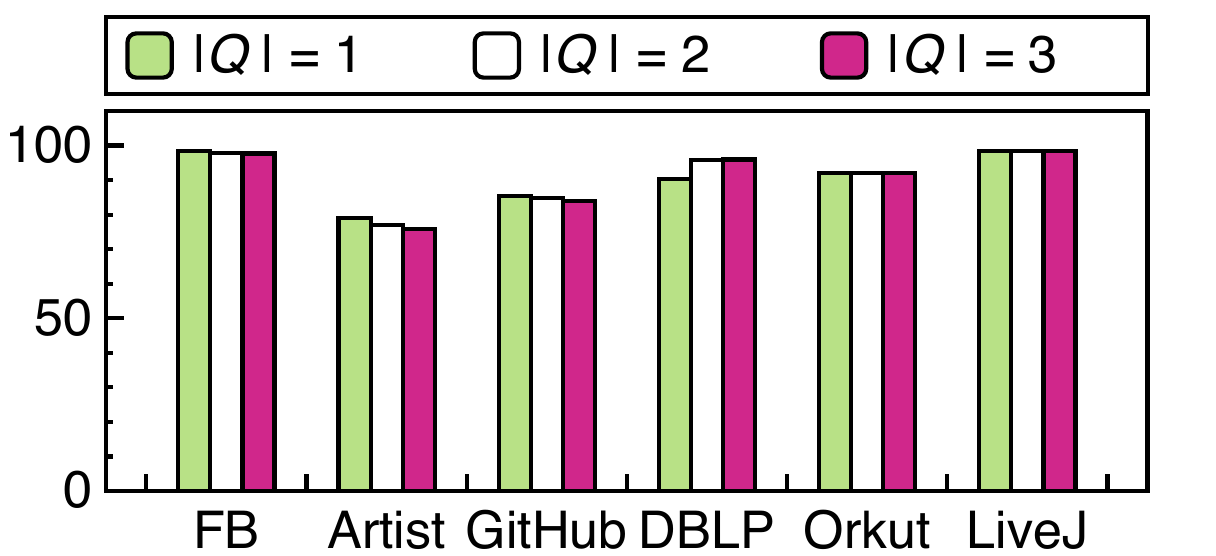}
    \vspace{-0.3cm}
  \end{minipage}
  }
  \hspace{0.1cm}
  \subfigure[Effect on runtime (ms)]{
  \begin{minipage}{0.45\linewidth}
    \includegraphics[scale=0.36]{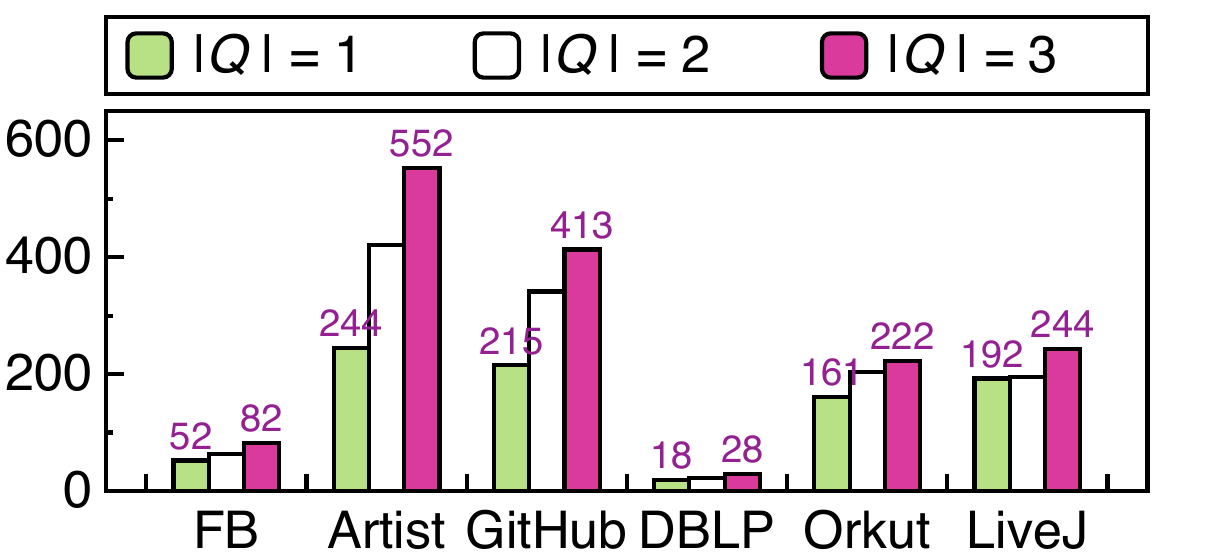}
    \vspace{-0.3cm}
  \end{minipage}
  }
  \caption{Effect of the \# query nodes $|Q|$}
  \label{fig:Q}
\end{figure}

	Figure \ref{fig:parameter}-\ref{fig:Q} show the parameter sensitivity for RW-TB.
	
	\vspace{0.1cm}
	\noindent\textbf{Effect of $m$.} In Figure \ref{fig:parameter}(a), the precision increases as $m$ increases. This is because the larger the $m$, the more the key-members are included in the $m$-bounded subgraph. The improvement gets stable after $m=2$, as most of key-members have been included in the 2-bounded subgraph and few key-members would be introduced by continuously increasing $m$. The runtime increases as $m$ increases as more time is required for random walk on a larger $m$-bounded subgraph.
	
	
	\vspace{0.1cm}
	\noindent\textbf{Effect of $r$.} In Figure \ref{fig:parameter}(c), the precision increases as $r$ increases and tends to be stable after 150 iterations. This is because the random walk nearly converges after 150 iterations so that has little effect on CKS's effectiveness. The runtime increases as $r$ increases (Figure \ref{fig:parameter}(d)), because the runtime is dominated by matrix multiplication. The more the iterations, the more the time is required for matrix multiplication. Finally, a trade-off can be achieved around $r=150$.
	
	\vspace{0.1cm}
	\noindent\textbf{Effect of $\alpha$.} We only study $\alpha$'s effect on CKS's effectiveness, because the runtime is dominated by random walk's efficiency (related to $m$ and $r$). The larger the $\alpha$, the more the average support increases or decreases. This would result in over-adjustment for average support, e.g., decrease (increase) an inflated (deflated) average support to an deflated (inflated) one, thus affecting the precision. If we set a small $\alpha$, then the amplitude of the fine-turning is too small to adjust average support to an appropriate value, thus affecting the precision. So, a moderately sized $\alpha$ is good for CKS, e.g., $\alpha=1$ for Facebook, DBLP, and LiveJournal, and $\alpha=1.5$ for others.
	
	
	
	\vspace{0.1cm}
	\noindent\textbf{Effect of $n$.} Since the size of ground truth (i.e., $|$ground truth$|$) for different queries is quite different, e.g., we may find a dozen key-members for some queries, while dozens of key-members for others, we set $n$ be a fraction $f\%$ of $|$ground truth$|$ ($f\%\in[10,100]$ in X axis). As Figure \ref{fig:topn} shows, precision is stable as $f\%$ increases. This is because most of the key-members have a large stationary visiting probabilities after random walk converges, so they can be successfully observed within the top-$n$ results. The recall increases as $f\%$ increases because we can find more key-members for a large $n$. As a result, $F1$ increases as $f\%$ increases. 
	
	
	\vspace{0.1cm}
	\noindent\textbf{Effect of $|Q|$.} Figure \ref{fig:Q}(b) shows that the more the query nodes, the larger the size of $m$-bounded subgraph of $Q$, leading more time for random walk. We always can find accurate enough top-$n$ key-members, as our random walk is performed based on the cohesiveness-aware transition matrix that can guide the random walk towards nodes of a cohesive $k$-truss.

	\subsection{Effect of Lightweight Refinement Method}
	\label{sec:refine}
	Table \ref{tab:refinement} shows the extra precision improvement and runtime increment by applying our refinement method with 1 and 2 iterations after RW-TB. The original precision is improved by 4.19\% and 5.20\% in total after 1 iteration and 2 iteration on average, but only introduce additional 2 ms in runtime.
	
	\section{Related Work}
	\label{sec:related_work}
\noindent\textbf{Truss decomposition algorithms.} Many cohesive subgraph models are studied to revealing potential community structures of real-world graphs, such as $k$-core \cite{Cheng2011,Khaouid2015}, $k$-truss \cite{Cohen2008,Huang2015}, $k$-ECC \cite{Chang2013,Zhou2012}, $k$-plex \cite{Conte2018}, and clique or quasi-clique \cite{Cui2013,Tsourakakis2013}. $k$-truss has been demonstrated to be an outstanding one as it achieves both high cohesiveness and high efficiency. \cite{Cohen2008} presents the first truss decomposition algorithm that can be kept in the main memory of a single machine, but cannot support large graphs. \cite{Wang2012,Zhao2012} propose I/O efficient truss decomposition for large graphs. Recently, many distributed truss computation algorithms have been developed \cite{Kabir2017,Cohen2009,Chen2014}, e.g., \cite{Cohen2009} proposes a distributed truss decomposition based on MapReduce. Moreover, to accelerate in-memory truss decomposition on billion-edge graphs, \cite{Che2020} propose a series of optimizations, e.g., intermediate results compacting and parallelizing on both multicore CPU and GPU.
	
	\vspace{0.1cm}
	\noindent\textbf{Truss-based community search (CS).} Since $k$-truss exhibits an inclusive hierarchy representing cores of a graph at different levels of granularity \cite{Akbas2017} and has some nice properties, e.g., a $k$-truss is diameter-bounded \cite{Huang2015}, it's usually adopted as the community model for CS \cite{Akbas2017,Huang2014,Huang2015,Liu2020a,Jiang2021}. The trussness is usually used for building trussness-based index or performing prunes, so the aforementioned truss decomposition algorithms are often invoked offline to compute trussness in advance, thus introducing extra overhead. Different from CS problem, we turn our perspective to the key-members in the community containing $q$ instead of the entire community. This motivates the CKS problem tackled in this paper.
	
	
	\vspace{0.1cm}
	\noindent\textbf{Critical node identification in complex networks.} This topic is related to our CKS problem. The difference comes from the definition of the critical nodes. Influence maximization is one widely used metric to find the critical nodes \cite{Zhu2020,Zhao2019,Wang2017,Ke2018}, they aim to find those nodes having the largest influence spread. Betweenness centrality (BC) computes the importance of a node in terms of total number of shortest paths passing through it \cite{Mumtaz2017,Freeman1977,Feng2022}. \cite{Zhang2017a,Zhang2018a,Zhang2018b} define the key-members of a network as those nodes that will cause a community collapse if they were removed, called collapsers. These critical node definitions are not optimal because they lack consideration of the close relation that naturally exists among key-members. Essentially, key-members in a network usually form a cohesive subgraph \cite{Calderoni2017,Liu2012}. This inspires us to define a $k$-truss based key-members in the CKS problem.
	

	\begin{table}
		\setlength{\abovecaptionskip}{0cm}
		\centering
		\caption{Effect of refinement on precision and runtime (ms)}
		\label{tab:refinement}
		\scalebox{0.6}{
			\begin{tabular}{c||cc||cc||cc||cc||cc||cc}
				\multirow{2}{*}{\textbf{Methods} $\downarrow$}       & \multicolumn{2}{c||}{Facebook}          & \multicolumn{2}{c||}{GitHub}            & \multicolumn{2}{c||}{Artist}            & \multicolumn{2}{c}{DBLP}		& \multicolumn{2}{c}{Orkut}	& \multicolumn{2}{c}{LiveJournal}              \\
				& \multicolumn{1}{c|}{$P$} & $T$  & \multicolumn{1}{c|}{$P$} & $T$  & \multicolumn{1}{c|}{$P$} & $T$  & \multicolumn{1}{c|}{$P$} & $T$ & \multicolumn{1}{c|}{$P$} & $T$  & \multicolumn{1}{c|}{$P$} & $T$ \\ 
				\hline\hline
				RW-TB & \multicolumn{1}{c|}{98.56}   & 52 & \multicolumn{1}{c|}{89.29}   & 215 & \multicolumn{1}{c|}{79.07}   & 244 & \multicolumn{1}{c|}{92.63}   & 18  & \multicolumn{1}{c|}{93.86}   & 161  & \multicolumn{1}{c|}{98.31}   & 192  \\ \hline
				RF $\times$ 1     & \multicolumn{1}{c|}{+0.72}   & +2  & \multicolumn{1}{c|}{+6.33}   & +1  & \multicolumn{1}{c|}{+6.61}   & +1  & \multicolumn{1}{c|}{+6.26}   & +1  & \multicolumn{1}{c|}{+4.30}   & +1 & \multicolumn{1}{c|}{+0.93}   & +4 \\ \hline
				RF $\times$ 2     & \multicolumn{1}{c|}{+0}        & +1  & \multicolumn{1}{c|}{+3.67}   & +1  & \multicolumn{1}{c|}{+2.17}   & +1  & \multicolumn{1}{c|}{+0}        & +1  & \multicolumn{1}{c|}{+0.23}   & +1 & \multicolumn{1}{c|}{+0}   & +1 \\
			\end{tabular}
		}
	\end{table}
	
	\section{Conclusion}
	\label{sec:conclude}
	We study the CKS problem that aims to seek the key-members of a cohesive community containing the query node. We first propose several exact algorithms atop an exact framework. Then, we present four random walk-based algorithms with several optimizations, by carefully considering some important cohesiveness features in the design of transition matrix. We theoretically analyze the rationality of designing cohesiveness-aware transition matrix, through Bayesian theory. Moreover, we propose a lightweight refinement method to refine the result and extend it for multiple query nodes. Extensive experiments demonstrate the superiority of our solution. In the future, we will try to extend our methods with more cohesive models, e.g., $k$-core, ego network with strict constraint, for heterogeneous, attributed, and uncertain graphs. 
	
	\vspace{-0.1cm}
	\section*{Acknowledgment}
	\vspace{-0.1cm}
		This work was supported by the National NSF of China (62072149 and 62006040), the Primary R\&D Plan of Zhejiang (2021C03156 and 2023C03198), and the Fundamental Research Funds for the Provincial Universities of Zhejiang (GK219909299001-006). We would like to express thanks to Key Laboratory of Brain Machine Collaborative Intelligence of Zhejiang Province (2020E10010).
	

\vspace{-0.2cm}
\begin{appendices}

\section{Analysis of the Effect of Query node Selection on CKS's Effectiveness}
\label{sec:analysis}
Given a query node $q$ and a key-member $u$, w.l.o.g., $q$ can visit $u$ via an edge (1-hop path) and other multi-hop paths. We say $q$ is structurally close to $u$ if $e_{qu}$ has a large trussness or the multi-hop path between them is short. Intuitively, the structurally closer $q$ is to $u$, the more probability that we can visit $u$ from $q$ during the random walk. This is also in line with some practical cases, e.g., it's always easier for the police to investigate a gang's key-members from their confidants than from an estranged suspect. In order to theoretically study the effect of the query node's structure closeness to key-members on our solution's effectiveness, we establish a concise Markov Chain model based on a hypergraph $H$ defined as follows.

\vspace{0.1cm}
\begin{myDef}
\label{def:hypergraph}
\textbf{Hypergraph $H$ over $G$}. Given a graph $G=(V_G,E_G)$ and a group of query nodes $Q\subseteq V_G$, we define the hypergraph over $G$ as $H=(V_H,E_H)$. (1) $V_H=V_G$ is the node set of $H$ that contains all the nodes from $G$, which is divided into three categories: query nodes $Q$, key-members $K$ w.r.t. $Q$, and $X=V_G\setminus \{Q\cup K\}$. (2) $E_H=\{e^H_{QK},e^H_{QX},e^H_{XK}\}$ is the hyperedge set of which each hyperedge connects two categories of nodes in $V_H$. (3) We assign a weight on each hyperedge as the largest trussness of an edge $e_{uv}\in E_G$ (Eq. \ref{eq:hypergraph}), where $u,v$ belong to the two categories connected by this hyperedge, denoted by $\phi(e^H_{ij})$ for $i,j\in\{Q,K,X\}$. For simplicity, we use $\gamma,\mu,\beta$ to indicate the weights on hyperedges $e^H_{QK},e^H_{QX},e^H_{XK}$, respectively.
\end{myDef}

\vspace{-0.1cm}
\begin{equation}
\small
\begin{split}
\label{eq:hypergraph}
\gamma=\phi(e^H_{QK})=\max\{\phi(e_{uv}): e_{uv}\in E_G, u\in Q, v\in K\}\\
\mu=\phi(e^H_{QX})=\max\{\phi(e_{uv}): e_{uv}\in E_G, u\in Q, v\in X\}\\
\beta=\phi(e^H_{XK})=\max\{\phi(e_{uv}): e_{uv}\in E_G, u\in X, v\in K\}
\end{split}
\end{equation}

\begin{figure}
\setlength{\abovecaptionskip}{-0.1cm}
\centering
\subfigcapskip=-0.18cm
\subfigure[Original model]{
\captionsetup{skip=0pt}
\includegraphics[scale=0.37]{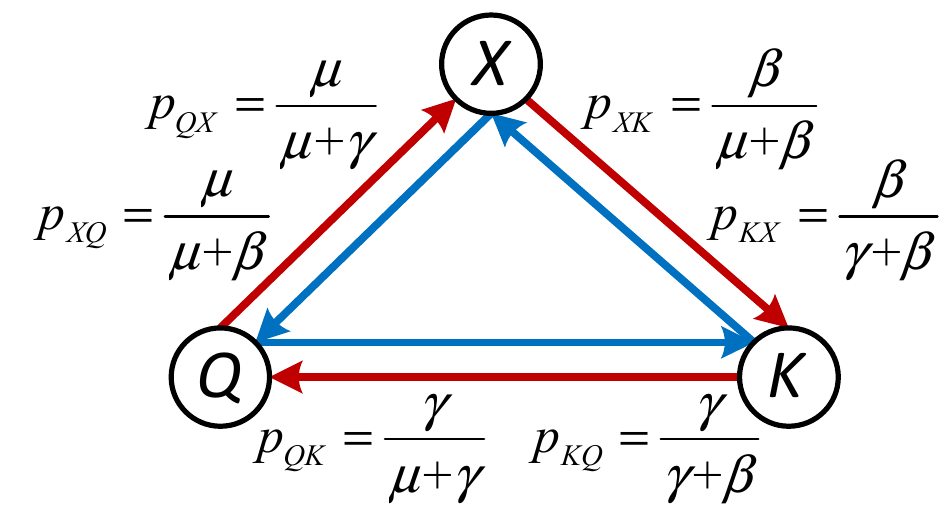}
\label{fig:model1}
}
\subfigure[Extended model]{
\captionsetup{skip=0pt}
\vspace{0mm}
\includegraphics[scale=0.37]{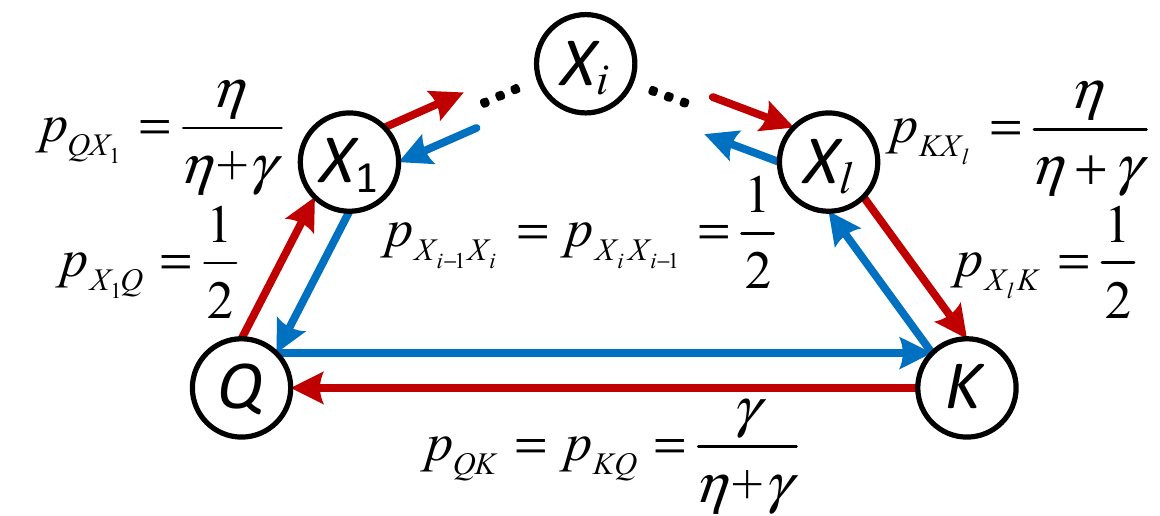}
\label{fig:model2}
}
\caption{Two Markov Chain models atop the hypergraph $H$}
\label{fig:rw_effect}
\end{figure}

Given a hypergraph $H$ over $G$, we can streamline the original Markov Chain over $G$ to a simple one over $H$. We initialize the transition matrix $\bm{P}$ via Eq. \ref{eq:transition1} by replacing the $sup(\cdot)$ with the weight on each hyperedge. Figure \ref{fig:model1} shows the concise model with transition probabilities. Notice that, the node $Q$ can visit key-members through a hyperedge $e^H_{QK}$ directly or a two-hop path indirectly. By dividing $X$ into several nodes $\{X_1,\cdots,X_l\}$, we can get a new model (discussed later) with a $(l+1)$-hop path from $Q$ to $K$, as shown in Figure \ref{fig:model2}. For the first model, we aim to study the effect of $\gamma$, i.e., the trussness of an edge (1-hop path), on $K$'s stationary visiting probability by given a specific $\mu$ and $\beta$. While for the second model, we try to study the effect of a multi-hop path's length on $K$'s stationary visiting probability by given a specific $\gamma$. 

Recall the first model, we apply \textit{matrix diagonalization} to get the expression of $\bm{\pi}^{(r)}$ after $r$ iterations as Eq. \ref{eq:diag}, where $\bm{A}$ is a invertible matrix and $\bm{D}$ is a diagonal matrix satisfying $\bm{P}=\bm{A}\times \bm{D}\times \bm{A}^{-1}$, and $\bm{D}^r$ means that $\bm{D}$ is multiplied $r$ times by itself.
\begin{equation}
\label{eq:diag}
\bm{\pi}^{(r)}=\bm{\pi}^{(0)}\times\bm{A}\times\bm{D}^r\times\bm{A}^{-1}
\end{equation}

\begin{myLemma}
\label{lemma:eigenvalue}
The eigenvalues $\lambda$ of the diagonal matrix $\bm{D}$ are solutions of the equation $(1-\lambda)[\lambda^2+\lambda+\frac{2\mu\beta\gamma}{(\mu+\beta)(\beta+\gamma)(\gamma+\mu)}]$=$0$.
\end{myLemma}

\vspace{0.1cm}
\begin{IEEEproof}
\label{pf:eigenvalue}
Given a specific $\bm{P}$ and an identity matrix $\bm{E}$, the eigenvalues of $\bm{D}$ is the solution of the following characteristic equation.
\end{IEEEproof}
\begin{equation}\nonumber
\begin{aligned}
\label{eq:charequation}
|\bm{P}-\lambda \bm{E}| & = \begin{bmatrix}
-\lambda   & \frac{\mu}{\mu +\gamma} & \frac{\gamma}{\mu +\gamma}\\
\frac{\mu}{\mu +\beta}   & -\lambda & \frac{\beta}{\mu +\beta}\\
\frac{\gamma}{\gamma +\beta}   & \frac{\beta}{\gamma +\beta} & -\lambda
\end{bmatrix} = (1-\lambda ) \begin{bmatrix}
1  & \frac{\mu}{\mu +\gamma} & \frac{\gamma}{\mu +\gamma}\\
1   & -\lambda & \frac{\beta}{\mu +\beta}\\
1   & \frac{\beta}{\gamma +\beta} & -\lambda
\end{bmatrix}\\
& = (1-\lambda)[\lambda^2+\lambda+\frac{2\mu\beta\gamma}{(\mu+\beta)(\beta+\gamma)(\gamma+\mu)}]=0
\end{aligned}
\end{equation}

Given the diagonal matrix $\bm{D}$, we can derive the expression of $\bm{\pi}^{(r)}$ by Eq. \ref{eq:diag}. Next, we leverage the real-world datasets to generate a set of instances of the first model, i.e., a set of instances of $H$, by randomly selecting $Q$ with different size. Each instance has a specific $\mu$, $\beta$, and $\gamma$, so we can directly substitute them into the expression of $\bm{\pi}^{(r)}$ to get $K$'s stationary visiting probability. Figure \ref{fig:rw_effect}(a-b) shows the effect of $\gamma$ for the following two cases.
\begin{itemize}[leftmargin=*]
\item \textbf{Case 1}. If $\mu,\beta>\gamma$, then we say that $Q$ is not structurally close to $K$. $K$'s stationary visiting probability in this case is smaller than that of $\mu,\beta<\gamma$ (Case 2). Moreover, the result for $\beta>\mu$ is better than that of $\beta<\mu$, indicating that our solution can visit key-members via a path in the direction of increasing trussness.
\item \textbf{Case 2}. If $\mu,\beta<\gamma$, then we say that $Q$ is structurally close to $K$. The result is better than Case 1. Besides, the result for $\beta>\mu$ is better than the opposite, indicating that a path in the direction of increasing trussness is helpful to improve the visiting probability.
\end{itemize}

\begin{figure}
\setlength{\abovecaptionskip}{-0.2cm}
\centering
\subfigcapskip=-0.18cm
\subfigure{
\captionsetup{skip=0pt}
\includegraphics[scale=0.8]{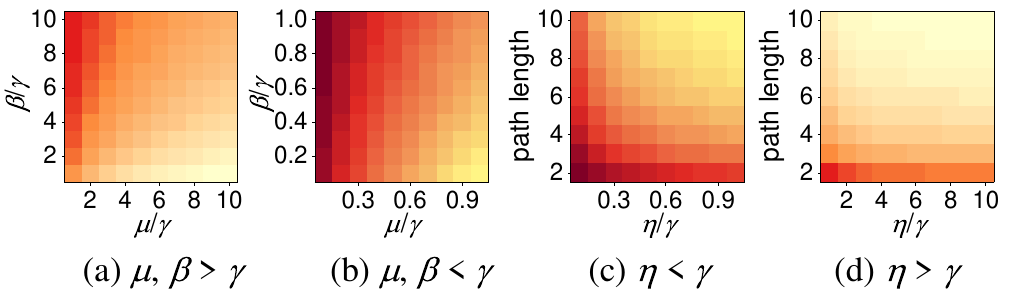}
}
\caption{$K$'s stationary visiting probability for two models: (a-b) original model and (c-d) extended model}
\label{fig:rw_effect}
\end{figure}

To confirm this experimentally, we perform our random walk-based algorithm mentioned in \S \ref{sec:rw_op3} for 1000 queries over original graph $G$ of real-world datasets, and then we compute the Pearson Correlation Coefficient (PCC) between the largest trussness $\gamma$ and the precision of top-$n$ key-members. Results are provided in Table \ref{tab:PCC}. For example, in \texttt{Artist}, it is medium positive correlated with precision (PCC of 0.41). We also adopt $t$-test \cite{Ugoni1995} to compute the significance of PCC, a value $<0.01$ indicates that this correlation is significant. Besides, we use the extended model to study the effect of the length of a multi-hop path between $Q$ and $K$ on the stationary visiting probability of $K$. For simplicity, we ignore the trussness difference of all edges on the $(l+1)$-hop path between $Q$ and $K$, and we set all edges on it as the same trussness, that is the largest one out of the $l+1$ edges, denoted by $\eta$ (Eq. \ref{eq:hypergraph1}).
\begin{equation}
\label{eq:hypergraph1}
\eta=\max\{\phi(e^H_{QX_1}),\phi(e^H_{X_lK}),\phi(e^H_{X_iX_{i+1}})|i\in[1,l-1]\}
\end{equation}

Next, we use the similar method to construct a set of instances of the extended model from the real-world datasets, and we show the effect of path length on the stationary visiting probability of $K$ given a specific relationship between $\gamma$ and $\eta$ in Figure \ref{fig:rw_effect}(c-d). Notice that, no matter what is the relationship between $\gamma$ and $\eta$, the path length has a negative effect on $K$'s visiting probability, that is the longer the path is, the smaller the visiting probability of $K$. From Table \ref{tab:PCC}, we see that the path length has PCC of -0.67 and significance of $<0.01$ for \texttt{Artist} dataset, indicating it has a strong negative correlation with the top-$n$ key-members' precision. 

\begin{table}[]
\setlength{\abovecaptionskip}{0cm}
\caption{PCC results w.r.t. CKS's precision}
\label{tab:PCC}
\scalebox{0.66}{
\begin{tabular}{c||cc||cc||cc||cc}
\multirow{2}{*}{\textbf{Factors} $\downarrow$}      & \multicolumn{2}{c||}{\textbf{Artist}}           & \multicolumn{2}{c||}{\textbf{Facebook}}             & \multicolumn{2}{c||}{\textbf{GitHub}}             & \multicolumn{2}{c}{\textbf{DBLP}}                \\
      & \multicolumn{1}{c|}{PCC} & Sig. & \multicolumn{1}{c|}{PCC} & Sig. & \multicolumn{1}{c|}{PCC} & Sig. & \multicolumn{1}{c|}{PCC} & Sig. \\ \hline\hline
Max trussness & \multicolumn{1}{c|}{0.41}      & $<0.01$   & \multicolumn{1}{c|}{0.07}      & 0.1   & \multicolumn{1}{c|}{0.24}      & $<0.01$   & \multicolumn{1}{c|}{0.17}      & $<0.01$   \\ \hline
Path length    & \multicolumn{1}{c|}{-0.67}     & $<0.01$   & \multicolumn{1}{c|}{-0.76}     & $<0.01$    & \multicolumn{1}{c|}{-0.73}      & $<0.01$ & \multicolumn{1}{c|}{-0.81}      & $<0.01$  
\end{tabular}
}
\end{table}	

\begin{table}[]
\setlength{\abovecaptionskip}{0cm}
\caption{Detailed precision (\%) w.r.t. maximum trussness and path length between query node and key-members}
\label{tab:detailed_precision}
\scalebox{0.86}{
\begin{tabular}{cccc||cccc}
\multicolumn{4}{c||}{\textbf{GitHub}}                                                                             & \multicolumn{4}{c}{\textbf{DBLP}}                                                                                \\ \hline \hline
\multicolumn{1}{c|}{Max $\phi$} & \multicolumn{1}{c|}{$P\%$}   & \multicolumn{1}{c|}{Hop} & $P\%$   & \multicolumn{1}{c||}{Max $\phi$} & \multicolumn{1}{c|}{$P\%$}   & \multicolumn{1}{c|}{Hop} & $P\%$   \\ \hline
\multicolumn{1}{c|}{2-4}        & \multicolumn{1}{c|}{79.48} & \multicolumn{1}{c|}{1-2}         & 87.63 & \multicolumn{1}{c||}{2-23}       & \multicolumn{1}{c|}{89.63} & \multicolumn{1}{c|}{1-4}         & 98.17 \\ \hline
\multicolumn{1}{c|}{5-7}        & \multicolumn{1}{c|}{70.88} & \multicolumn{1}{c|}{3-4}         & 82.14 & \multicolumn{1}{c||}{24-45}      & \multicolumn{1}{c|}{100}   & \multicolumn{1}{c|}{5-7}         & 93.50 \\ \hline
\multicolumn{1}{c|}{8-10}       & \multicolumn{1}{c|}{83.54} & \multicolumn{1}{c|}{5-7}         & 14.75 & \multicolumn{1}{c||}{46-65}      & \multicolumn{1}{c|}{100}   & \multicolumn{1}{c|}{8-11}        & 22.99
\end{tabular}
}
\end{table}

\vspace{0.1cm}
\noindent\textbf{Effect of maximum trussness and path length on real-world datasets.} We provide the detailed precision results w.r.t. the maximum trussness and shortest path length between the query node and key-members. As shown in Table \ref{tab:detailed_precision}, we divide the precision results into several intervals of the maximum trussness and path length to see their effect on effectiveness. Since different datasets have different characteristics, the interval setting is also different. We found that the larger the maximum trussness and the shorter the multi-hop path, the higher the precision. This experimentally proves that the accuracy analysis in \S \ref{sec:analysis} is correct.

\end{appendices}

	\bibliographystyle{IEEEtran}
	\bibliography{CS}

\begin{thebibliography}{10}
\providecommand{\url}[1]{#1}
\csname url@samestyle\endcsname
\providecommand{\newblock}{\relax}
\providecommand{\bibinfo}[2]{#2}
\providecommand{\BIBentrySTDinterwordspacing}{\spaceskip=0pt\relax}
\providecommand{\BIBentryALTinterwordstretchfactor}{4}
\providecommand{\BIBentryALTinterwordspacing}{\spaceskip=\fontdimen2\font plus
\BIBentryALTinterwordstretchfactor\fontdimen3\font minus
  \fontdimen4\font\relax}
\providecommand{\BIBforeignlanguage}[2]{{%
\expandafter\ifx\csname l@#1\endcsname\relax
\typeout{** WARNING: IEEEtran.bst: No hyphenation pattern has been}%
\typeout{** loaded for the language `#1'. Using the pattern for}%
\typeout{** the default language instead.}%
\else
\language=\csname l@#1\endcsname
\fi
#2}}
\providecommand{\BIBdecl}{\relax}
\BIBdecl

\bibitem{Zhang2019}
Z.~Zhang, X.~Huang, J.~Xu, B.~Choi, and Z.~Shang, ``Keyword-centric community
  search,'' in \emph{ICDE}, 2019, pp. 422--433.

\bibitem{Fang2017}
Y.~Fang, R.~Cheng, X.~Li, S.~Luo, and J.~Hu, ``Effective {C}ommunity {S}earch
  over {L}arge {S}patial {G}raphs,'' \emph{PVLDB}, vol.~10, no.~6, pp.
  709--720, 2017.

\bibitem{Huang2017}
X.~Huang and L.~V.~S. Lakshmanan, ``Attribute-driven community search,''
  \emph{PVLDB}, vol.~10, no.~9, pp. 949--960, 2017.

\bibitem{Sun2020}
L.~Sun, X.~Huang, R.~Li, B.~Choi, and J.~Xu, ``Index-based intimate-core
  community search in large weighted graphs,'' \emph{{IEEE} Trans. Knowl. Data
  Eng.}, 2020.

\bibitem{Liu2020}
Q.~Liu, Y.~Zhu, M.~Zhao, X.~Huang, J.~Xu, and Y.~Gao, ``{VAC:} vertex-centric
  attributed community search,'' in \emph{ICDE}, 2020, pp. 937--948.

\bibitem{Sozio2010}
M.~Sozio and A.~Gionis, ``The community-search problem and how to plan a
  successful cocktail party,'' in \emph{KDD}, 2010, pp. 939--948.

\bibitem{Xu2022}
X.~Xu, J.~Liu, Y.~Wang, and X.~Ke, ``Academic expert finding via (k,{P})-core
  based embedding over heterogeneous graphs,'' in \emph{ICDE}, 2022, pp.
  338--351.

\bibitem{Wang2023}
Y.~Wang, J.~Liu, X.~Xu, X.~Ke, T.~Wu, and X.~Gou, ``Efficient and effective
  academic expert finding on heterogeneous graphs through (k,{P})-core based
  embedding,'' \emph{{ACM} Trans. Knowl. Discov. Data}, vol.~17, no.~6, pp.
  85:1--85:35, 2023.

\bibitem{Cui2014}
W.~Cui, Y.~Xiao, H.~Wang, and W.~Wang, ``Local {S}earch of {C}ommunities in
  {L}arge {G}raphs,'' in \emph{{SIGMOD}}, 2014, pp. 991--1002.

\bibitem{Huang2015}
X.~Huang, L.~V.~S. Lakshmanan, J.~X. Yu, and H.~Cheng, ``Approximate {C}losest
  {C}ommunity {S}earch in {N}etworks,'' \emph{PVLDB}, vol.~9, no.~4, pp.
  276--287, 2015.

\bibitem{Hu2016}
J.~Hu, X.~Wu, R.~Cheng, S.~Luo, and Y.~Fang, ``Querying {M}inimal {S}teiner
  {M}aximum-connected {S}ubgraphs in {L}arge {G}raphs,'' in \emph{CIKM}, 2016,
  pp. 1241--1250.

\bibitem{Cui2013}
W.~Cui, Y.~Xiao, H.~Wang, Y.~Lu, and W.~Wang, ``Online {S}earch of
  {O}verlapping {C}ommunities,'' in \emph{SIGMOD}, 2013, pp. 277--288.

\bibitem{Yao2021}
K.~Yao and L.~Chang, ``Efficient size-bounded community search over large
  networks,'' \emph{PVLDB}, vol.~14, no.~8, pp. 1441--1453, 2021.

\bibitem{Liu2021}
B.~Liu, F.~Zhang, W.~Zhang, X.~Lin, and Y.~Zhang, ``Efficient community search
  with size constraint,'' in \emph{ICDE}, 2021, pp. 97--108.

\bibitem{Kim2022}
J.~Kim, S.~Luo, G.~Cong, and W.~Yu, ``{DMCS} : Density modularity based
  community search,'' in \emph{SIGMOD}, 2022, pp. 889--903.

\bibitem{Rozemberczki2019}
B.~Rozemberczki, R.~Davies, R.~Sarkar, and C.~Sutton, ``{GEMSEC:} graph
  embedding with self clustering,'' in \emph{{ASONAM}}, 2019, pp. 65--72.

\bibitem{Liu2022}
Y.~Liu, A.~Song, X.~Shan, Y.~Xue, and J.~Jin, ``Identifying critical nodes in
  power networks: {A} group-driven framework,'' \emph{Expert Syst. Appl.}, vol.
  196, p. 116557, 2022.

\bibitem{Munikoti2022}
S.~Munikoti, L.~Das, and B.~Natarajan, ``Scalable graph neural network-based
  framework for identifying critical nodes and links in complex networks,''
  \emph{Neurocomputing}, vol. 468, pp. 211--221, 2022.

\bibitem{Sun2022}
R.~Sun, C.~Chen, X.~Liu, S.~Xu, X.~Wang, and X.~Lin, ``Critical nodes
  identification in large networks: The inclined and detached models,''
  \emph{World Wide Web}, vol.~25, no.~3, pp. 1315--1341, 2022.

\bibitem{Xu2019}
M.~Xu, J.~Wu, M.~Liu, Y.~Xiao, H.~Wang, and D.~Hu, ``Discovery of critical
  nodes in road networks through mining from vehicle trajectories,''
  \emph{{IEEE} Trans. Intell. Transp. Syst.}, vol.~20, no.~2, pp. 583--593,
  2019.

\bibitem{Hepenstal2021}
S.~Hepenstal, L.~Zhang, N.~Kodagoda, and B.~L.~W. Wong, ``Developing
  conversational agents for use in criminal investigations,'' \emph{{ACM}
  Trans. Interact. Intell. Syst.}, vol.~11, no. 3-4, pp. 25:1--25:35, 2021.

\bibitem{Calderoni2017}
F.~Calderoni, D.~Brunetto, and C.~Piccardi, ``Communities in criminal networks:
  {A} case study,'' \emph{Soc. Networks}, vol.~48, pp. 116--125, 2017.

\bibitem{Liu2012}
X.~Liu, E.~Patacchini, Y.~Zenou, and L.~fei Lee, ``Criminal networks: Who is
  the key player,'' \emph{CEPR Discussion Paper}, 2012.

\bibitem{Zhu2020}
Y.~Zhu, J.~Tang, and X.~Tang, ``Pricing influential nodes in online social
  networks,'' \emph{PVLDB}, vol.~13, no.~10, pp. 1614--1627, 2020.

\bibitem{Zhao2019}
J.~Zhao, S.~Shang, P.~Wang, J.~C.~S. Lui, and X.~Zhang, ``Tracking influential
  nodes in time-decaying dynamic interaction networks,'' in \emph{ICDE}, 2019,
  pp. 1106--1117.

\bibitem{Wang2017}
X.~Wang, Y.~Zhang, W.~Zhang, X.~Lin, and C.~Chen, ``Bring order into the
  samples: {A} novel scalable method for influence maximization (extended
  abstract),'' in \emph{ICDE}, 2017, pp. 55--56.

\bibitem{Mumtaz2017}
S.~Mumtaz and X.~Wang, ``Identifying top-k influential nodes in networks,'' in
  \emph{CIKM}, 2017, pp. 2219--2222.

\bibitem{Zhang2017a}
F.~Zhang, Y.~Zhang, L.~Qin, W.~Zhang, and X.~Lin, ``Finding critical users for
  social network engagement: The collapsed k-core problem,'' in \emph{AAAI},
  2017, pp. 245--251.

\bibitem{McAuley2014}
J.~J. McAuley and J.~Leskovec, ``Discovering social circles in ego networks,''
  \emph{{ACM} Trans. Knowl. Discov. Data}, vol.~8, no.~1, pp. 4:1--4:28, 2014.

\bibitem{Wang2012}
J.~Wang and J.~Cheng, ``Truss decomposition in massive networks,''
  \emph{PVLDB}, vol.~5, no.~9, pp. 812--823, 2012.

\bibitem{Che2020}
Y.~Che, Z.~Lai, S.~Sun, Y.~Wang, and Q.~Luo, ``Accelerating truss decomposition
  on heterogeneous processors,'' \emph{PVLDB}, vol.~13, no.~10, pp. 1751--1764,
  2020.

\bibitem{Jiang2021}
Y.~Jiang, X.~Huang, and H.~Cheng, ``{I/O} efficient k-truss community search in
  massive graphs,'' \emph{{VLDB} Journay}, vol.~30, no.~5, pp. 713--738, 2021.

\bibitem{Huang2014}
X.~Huang, H.~Cheng, L.~Qin, W.~Tian, and J.~X. Yu, ``Querying {K}-truss
  {C}ommunity in large and dynamic graphs,'' in \emph{SIGMOD}, 2014, pp.
  1311--1322.

\bibitem{Doane2011}
D.~P. Doane and L.~E. Seward, ``Measuring skewness: A forgotten statistic?''
  \emph{Journal of statistics education}, vol.~19, no.~2, 2011.

\bibitem{VonHippel2005}
P.~T. Von~Hippel, ``Mean, median, and skew: Correcting a textbook rule,''
  \emph{Journal of statistics Education}, vol.~13, no.~2, 2005.

\bibitem{Groeneveld1984}
R.~A. Groeneveld and G.~Meeden, ``Measuring skewness and kurtosis,''
  \emph{Journal of the Royal Statistical Society: Series D (The Statistician)},
  vol.~33, no.~4, pp. 391--399, 1984.

\bibitem{Yang2020}
Y.~Yang, Y.~Fang, X.~Lin, and W.~Zhang, ``{Effective and Efficient Truss
  Computation over Large Heterogeneous Information Networks},'' in \emph{ICDE},
  2020, pp. 901--912.

\bibitem{Liu2020a}
Q.~Liu, M.~Zhao, X.~Huang, J.~Xu, and Y.~Gao, ``Truss-based community search
  over large directed graphs,'' in \emph{SIGMOD}, 2020, pp. 2183--2197.

\bibitem{Zhao2015}
J.~Zhao, J.~C. Lui, D.~Towsley, P.~Wang, and X.~Guan, ``A {T}ale of {T}hree
  {G}raphs: {S}ampling {D}esign on {H}ybrid {S}ocial-{A}ffiliation
  {N}etworks,'' in \emph{ICDE}, 2015.

\bibitem{Li2019}
Y.~Li, Z.~Wu, S.~Lin, H.~Xie, M.~Lv, Y.~Xu, and J.~C.~S. Lui, ``Walking with
  perception: Efficient random walk sampling via common neighbor awareness,''
  in \emph{ICDE}, 2019, pp. 962--973.

\bibitem{Wang2022a}
Y.~Wang, A.~Khan, X.~Xu, J.~Jin, Q.~Hong, and T.~Fu, ``Aggregate queries on
  knowledge graphs: Fast approximation with semantic-aware sampling,'' in
  \emph{ICDE}, 2022, pp. 2914--2927.

\bibitem{Kleinberg2000}
J.~M. Kleinberg, ``Navigation in a small world,'' \emph{Nature}, vol. 406, pp.
  845--845, 2000.

\bibitem{Malkov2020}
Y.~A. Malkov and D.~A. Yashunin, ``Efficient and robust approximate nearest
  neighbor search using hierarchical navigable small world graphs,''
  \emph{{IEEE} Trans. Pattern Anal. Mach. Intell.}, vol.~42, no.~4, pp.
  824--836, 2020.

\bibitem{Yang2012}
S.~Yang, X.~Yan, B.~Zong, and A.~Khan, ``Towards effective partition management
  for large graphs,'' in \emph{SIGMOD}, 2012, pp. 517--528.

\bibitem{Ross2014}
S.~M. Ross, \emph{Introduction to {P}robability {M}odels}.\hskip 1em plus 0.5em
  minus 0.4em\relax Academic press, 2014.

\bibitem{skewness}
``{Skewness},'' \url{https://en.wikipedia.org/wiki/Skewness}, 2022.

\bibitem{Dean2018}
S.~Dean and B.~Illowsky, ``Descriptive statistics: Skewness and the mean,
  median, and mode,'' \emph{Connexions website}, 2018.

\bibitem{John2013}
G.~H. John and P.~Langley, ``Estimating continuous distributions in bayesian
  classifiers,'' \emph{arXiv}, vol. abs/1302.4964, 2013.

\bibitem{Sur2015}
\BIBentryALTinterwordspacing
P.~Sur, G.~Shmueli, S.~Bose, and P.~Dubey, ``Modeling bimodal discrete data
  using conway-maxwell-poisson mixture models,'' \emph{Journal of Business \&
  Economic Statistics}, vol.~33, no.~3, pp. 352--365, 2015. [Online].
  Available: \url{https://doi.org/10.1080/07350015.2014.949343}
\BIBentrySTDinterwordspacing

\bibitem{Velez2015}
J.~Vélez, J.~Correa, and F.~Marmolejo-Ramos, ``A new approach to the box–cox
  transformation,'' \emph{Frontiers in Applied Mathematics and Statistics},
  vol.~1, 10 2015.

\bibitem{Escarela2009}
G.~Escarela and A.~Hernandez, ``Modelling random couples using copulas,''
  \emph{Revista Colombiana de Estadística}, vol.~32, pp. 33--58, 06 2009.

\bibitem{Reynolds2015}
\BIBentryALTinterwordspacing
D.~Reynolds, \emph{Gaussian Mixture Models}.\hskip 1em plus 0.5em minus
  0.4em\relax Boston, MA: Springer US, 2015, pp. 827--832. [Online]. Available:
  \url{https://doi.org/10.1007/978-1-4899-7488-4_196}
\BIBentrySTDinterwordspacing

\bibitem{Box1964}
G.~Box and D.~Cox, ``An analysis of transformations.'' vol.~26, pp. 211--243,
  07 1964.

\bibitem{code}
O.~code and datasets, ``Code and datasets,''
  \url{https://github.com/KGLab-HDU/CKS}, 2023.

\bibitem{McAuley2012}
J.~J. McAuley and J.~Leskovec, ``Learning to discover social circles in ego
  networks,'' in \emph{NIPS}, 2012, pp. 548--556.

\bibitem{Rozemberczki2021}
B.~Rozemberczki, C.~Allen, and R.~Sarkar, ``Multi-scale attributed node
  embedding,'' \emph{J. Complex Networks}, vol.~9, no.~2, 2021.

\bibitem{Yang2015}
J.~Yang and J.~Leskovec, ``Defining and evaluating network communities based on
  ground-truth,'' \emph{Knowl. Inf. Syst.}, vol.~42, no.~1, pp. 181--213, 2015.

\bibitem{Orkut2022}
Orkut, ``Orkut,''
  \url{https://www.comp.hkbu.edu.hk/~db/book/community_search.html}, 2022.

\bibitem{LiveJournal2022}
LiveJournal, ``Livejournal,''
  \url{http://snap.stanford.edu/data/com-LiveJournal.html}, 2022.

\bibitem{Wu2015}
Y.~Wu, R.~Jin, J.~Li, and X.~Zhang, ``Robust local community detection: On free
  rider effect and its elimination,'' \emph{PVLDB}, vol.~8, no.~7, pp.
  798--809, 2015.

\bibitem{Cheng2011}
J.~Cheng, Y.~Ke, S.~Chu, and M.~T. {\"{O}}zsu, ``Efficient core decomposition
  in massive networks,'' in \emph{ICDE}, 2011, pp. 51--62.

\bibitem{Khaouid2015}
W.~Khaouid, M.~Barsky, S.~Venkatesh, and A.~Thomo, ``K-core decomposition of
  large networks on a single {PC},'' \emph{PVLDB}, vol.~9, no.~1, pp. 13--23,
  2015.

\bibitem{Cohen2008}
J.~Cohen, ``Trusses: Cohesive subgraphs for social network analysis,'' Tech.
  Rep., 2008.

\bibitem{Chang2013}
L.~Chang, J.~X. Yu, L.~Qin, X.~Lin, C.~Liu, and W.~Liang, ``Efficiently
  computing k-edge connected components via graph decomposition,'' in
  \emph{SIGMOD}, 2013, pp. 205--216.

\bibitem{Zhou2012}
R.~Zhou, C.~Liu, J.~X. Yu, W.~Liang, B.~Chen, and J.~Li, ``Finding maximal
  k-edge-connected subgraphs from a large graph,'' in \emph{EDBT}, 2012, pp.
  480--491.

\bibitem{Conte2018}
A.~Conte, T.~D. Matteis, D.~D. Sensi, R.~Grossi, A.~Marino, and L.~Versari,
  ``{D2K:} scalable community detection in massive networks via small-diameter
  k-plexes,'' in \emph{SIGKDD}, 2018, pp. 1272--1281.

\bibitem{Tsourakakis2013}
C.~E. Tsourakakis, F.~Bonchi, A.~Gionis, F.~Gullo, and M.~A. Tsiarli, ``Denser
  than the densest subgraph: Extracting optimal quasi-cliques with quality
  guarantees,'' in \emph{SIGKDD}, 2013, pp. 104--112.

\bibitem{Zhao2012}
F.~Zhao and A.~K.~H. Tung, ``Large scale cohesive subgraphs discovery for
  social network visual analysis,'' \emph{PVLDB}, vol.~6, no.~2, pp. 85--96,
  2012.

\bibitem{Kabir2017}
H.~Kabir and K.~Madduri, ``Shared-memory graph truss decomposition,'' in
  \emph{HIPC}, 2017, pp. 13--22.

\bibitem{Cohen2009}
J.~D. Cohen, ``Graph twiddling in a mapreduce world,'' \emph{Comput. Sci.
  Eng.}, vol.~11, no.~4, pp. 29--41, 2009.

\bibitem{Chen2014}
P.~Chen, C.~Chou, and M.~Chen, ``Distributed algorithms for k-truss
  decomposition,'' in \emph{BigData}, 2014, pp. 471--480.

\bibitem{Akbas2017}
E.~Akbas and P.~Zhao, ``Truss-based community search: a truss-equivalence based
  indexing approach,'' \emph{PVLDB}, vol.~10, no.~11, pp. 1298--1309, 2017.

\bibitem{Ke2018}
X.~Ke, A.~Khan, and G.~Cong, ``Finding seeds and relevant tags jointly: For
  targeted influence maximization in social networks,'' in \emph{SIGMOD}, 2018,
  pp. 1097--1111.

\bibitem{Freeman1977}
L.~Freeman, ``A set of measures of centrality based on betweenness,''
  \emph{Sociometry}, vol.~40, pp. 35--41, 03 1977.

\bibitem{Feng2022}
Y.~Feng, H.~Wang, and H.~Lu, ``A faster algorithm for betweenness centrality
  based on adjacency matrices,'' \emph{arXiv:2205.00162}, 2022.

\bibitem{Zhang2018a}
F.~Zhang, Y.~Zhang, L.~Qin, W.~Zhang, and X.~Lin, ``Efficiently reinforcing
  social networks over user engagement and tie strength,'' in \emph{ICDE},
  2018, pp. 557--568.

\bibitem{Zhang2018b}
F.~Zhang, L.~Yuan, Y.~Zhang, L.~Qin, X.~Lin, and A.~Zhou, ``Discovering strong
  communities with user engagement and tie strength,'' in \emph{DASFAA}, vol.
  10827, 2018, pp. 425--441.

\bibitem{Ugoni1995}
A.~Ugoni and B.~F. Walker, ``The t-test: An introduction,'' \emph{COMSIG
  review}, vol.~4, no.~2, p.~37, 1995.

\end{thebibliography}
	
\end{document}